\documentclass[%
reprint,
superscriptaddress,
mathtools,
aps, 
]{revtex4-1}

\usepackage{graphicx}
\graphicspath{{Figures/}}
\usepackage{epstopdf}
\usepackage{dcolumn}
\usepackage{bm}
\usepackage{comment}
\usepackage{color}
\usepackage{physics}
\usepackage{siunitx}
\usepackage{chemformula}
\usepackage{ragged2e}
\usepackage[flushleft]{threeparttable}
\usepackage{hyperref} 
\hypersetup{
    colorlinks  = true,
    citecolor   = blue,
    linkcolor   = blue,
    urlcolor   =blue
}
\usepackage[all]{hypcap}

\begin{document}

\title{Physical Mechanism behind the Hysteresis-free Negative Capacitance Effect in Metal-Ferroelectric-Insulator-Metal Capacitors with Dielectric Leakage and Interfacial Trapped Charges}

\author{Chia-Sheng Hsu}
\affiliation{School of Electrical and Computer Engineering, Georgia Institute of Technology, Atlanta, GA 30332, USA
}%
\email{chiasheng@gatech.edu}
\author{Sou-Chi Chang}%
\affiliation{Components Research, Intel Corporation, Hillsboro, OR 97124, USA}%
\author{Dmitri E. Nikonov}%
\affiliation{Components Research, Intel Corporation, Hillsboro, OR 97124, USA}%
\author{Ian A. Young}%
\affiliation{Components Research, Intel Corporation, Hillsboro, OR 97124, USA}%
\author{Azad Naeemi}%
\affiliation{School of Electrical and Computer Engineering, Georgia Institute of Technology, Atlanta, GA 30332, USA
}%


\begin{abstract}
The negative capacitance (NC) stabilization of a ferroelectric (FE) material can potentially provide an alternative way to further reduce the power consumption in ultra-scaled devices and thus has been of great interest in technology and science in the past decade. In this article, we present a physical picture for a better understanding of the hysteresis-free charge boost effect observed experimentally in metal-ferroelectric-insulator-metal (MFIM) capacitors. By introducing the dielectric (DE) leakage and interfacial trapped charges, our simulations of the hysteresis loops are in a strong agreement with the experimental measurements, suggesting the existence of an interfacial oxide layer at the FE-metal interface in metal-ferroelectric-metal (MFM) capacitors. Based on the pulse switching measurements, we find that the charge enhancement and hysteresis are dominated by the FE domain viscosity and DE leakage, respectively. Our simulation results show that the underlying mechanisms for the observed hysteresis-free charge enhancement in MFIM may be physically different from the alleged NC stabilization and capacitance matching. Moreover, the link between Merz's law and the phenomenological kinetic coefficient is discussed, and the possible cause of the residual charges observed after pulse switching is explained by the trapped charge dynamics at the FE-DE interface. The physical interpretation presented in this work can provide important insights into the NC effect in MFIM capacitors and future studies of low-power logic devices.


\end{abstract}

\pacs{Valid PACS appear here}
\maketitle


\section{\label{sec:level1}Introduction}
{A}{s} the relentless pursuit of device miniaturization goes into the nanometer regime, Moore's law has gradually come to a bottleneck due to the fact that the power dissipation in microchips becomes a more and more challenging concern~\cite{Moore1998,Kim2003}. Recently, a stack structure consisting of an FE layer and a DE layer was proposed to achieve voltage amplification in a DE layer~\cite{Salahuddin2008}. The physical concept behind this approach is that the metastable negative capacitance (NC) state arising from the double-well energy profile of the FE can be stabilized by the DE layer in terms of the total free energy of the system. Such a proposal may provide a potential way to significantly improve the subthreshold swing of conventional complementary metal-oxide-semiconductor (CMOS) transistors at room temperature~\cite{Chang2017a,Hsu2018}. \par

Ferroelectrics are materials that exhibit the properties: (i) the electric polarization can be reversed by an externally applied voltage and (ii) the remanent polarization remains nonvolatile under zero bias. These unique properties have made ferroelectrics promising materials for voltage-controlled nonvolatile memory devices~\cite{Ni2018a,Ni2018}. In the past decade since the proposal of using NC for low-power logic devices, FE-based capacitors, including MFM, MFIM and metal-ferroelectric-metal-insulator-metal (MFMIM), have been intensively investigated experimentally and theoretically in search of the evidence for the transient and steady-state NC effects~\cite{Khan2014,Gao2014,Appleby2014,Kim2016,Hoffmann2016,Chang2018,Hoffmann2018,Hoffmann2019,Kim2019,Hsu2020}. In particular, recently discovered doped hafnium oxides are widely used as the FE layer due to the high scalability and CMOS process compatibility~\cite{Boescke2011,Mueller2012,Sharma2017}. \par

With an MFM capacitor connected in series with a large resistor, the transient NC behaviors have been observed in various FE materials, including perovskite \ch{Pb(Zr_{0.2}Ti_{0.8})O3} and various doped hafnium oxides~\cite{Khan2014,Hoffmann2016,Kobayashi2016}. The physical origin of the observed transient NC lies in the mismatch between the switching rates of free charges and bound charges (FE polarization) in the resistor-capacitor circuit~\cite{Chang2018}. To further seek the evidence of static NC stabilization, FE-DE stacks are the key devices to be investigated according to the NC theory. Among MFIM, MFMIM and FE-DE superlattices, MFIM capacitors are of great importance because of the structure similarity to the ferroelectric field effect transistors (FeFETs). For MFIM stacks, it is unlikely to directly measure the internal DE voltage amplification, which was theoretically proposed to be achieved by the steady-state NC stabilization. Therefore, a considerable amount of experimental efforts have been focused on the evidence of capacitance enhancement in an MFIM capacitor compared to the associated standalone DE capacitor~\cite{Alam2019}. Such a capacitance enhancement is referred to as the charge boost effect. \par

In the recent research progress, it was experimentally observed that the charge boost effect and hysteresis-free static ``S-shaped curve" could be achieved in MFIM with short pulse measurements~\cite{Hoffmann2018,Hoffmann2019}. On one hand, the charge boost may indicate that the MFIM capacitor has a larger capacitance than the associated DE capacitor, which was allegedly caused by the NC stabilization. On the other hand, the observed hysteresis-free FE switching (S-curve), which is predicted by the Landau phenomenological formalism, is a required characteristic for the logic applications. However, the physical mechanisms for these observations are still not clear. For example, Liu \textit{et al.} have recently proposed an alternative perspective on such experimental observations~\cite{Liu2020}. Therefore, it is of great importance to further explore the underlying physical mechanisms for such prospective experimental evidence of NC stabilization. \par

In this paper, we establish a physical model for MFM and MFIM capacitors by introducing the inevitable DE leakage and trapped charges at the FE-DE interface. We show that the experimentally measured hysteresis loop of MFM capacitors can be well described by the Landau formalism with the proposed physical mechanisms included. The charge boost and hysteresis behaviors observed in the pulse measurements are found to be directly influenced by the FE intrinsic domain viscosity and DE leakage, respectively. The kinetic coefficients extracted from the charge responses are found to be linked to the well-known Merz's law~\cite{Merz1956} in the NC region. Our simulation results suggest that the experimentally observed hysteresis-free capacitance enhancement of MFIM may be caused not by the NC stabilization and capacitance matching but by the material properties of the heterostructure. Furthermore, the possible cause of the experimentally observed residual charges may be explained by the trapped charges existing at the FE-DE interface. \par

This paper is organized as follows. In Sec.~\ref{sec2}, the theoretical approach is presented to describe the switching characteristics of MFM and MFIM capacitors, with the DE leakage and trapped charge mechanisms included. In Sec.~\ref{sec3}, based on the physical model, the FE material parameters can be well extracted using the hysteresis measurements. More importantly, the experimentally observed charge boost and hysteresis-free static S-curve are well captured, and a clear physical picture for such phenomena is provided and discussed in detail. In Sec.~\ref{sec4}, we conclude this work by highlighting the underlying mechanisms for experimental observations.

\section{\label{sec:level1}Theoretical Formalism} \label{sec2}
\begin{figure}[!t]
\centering
\includegraphics[width=0.4\textwidth]{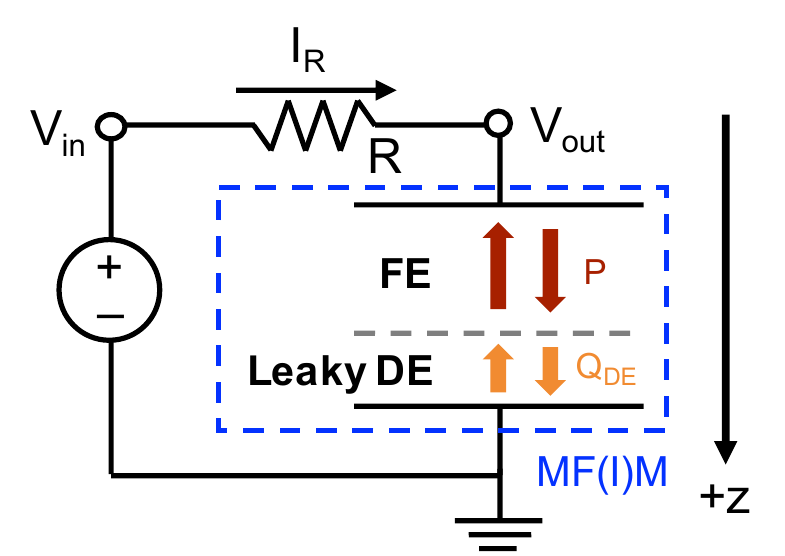}
\caption{The schematic diagram of a resistor-capacitor circuit for the study of switching dynamics in the experimental measurements. The positive axis is defined as the z direction toward ground.}
\label{fig:circuit}
\end{figure}

To describe the electrical properties of MFIM bilayer stacks measured in the experiments, Kirchhoff's law is applied for the schematic circuit diagram shown in Fig.~\ref{fig:circuit}. The total current $I_\text{R}$ flowing through the series resistor is thus expressed as
\begin{equation} \label{eq:KVL}
I_\text{R} = A\pdv{Q_\text{f}}{t} = \frac{V_\text{in} - V_\text{out}}{R},
\end{equation}
where ${Q_\text{f}}$ is the free charge density, $R$ is the series resistance and $A$ is the capacitor area. In addition, the following conditions have to be satisfied by assuming the electric displacement field is continuous at material boundaries:
\begin{equation} \label{eq:KCL}
I_\text{R} = I_\text{FE} = I_\text{DE} + I_\text{L},
\end{equation}
where $I_\text{L}$ represents the DE leakage current, and $I_\text{FE}$ and $I_\text{DE}$ are the displacement current of the FE and the DE layers, respectively. Instead of the conventional DC leakage current, $I_\text{L}$ here describes the transient DE leakage due to the voltage across the DE layer, as will be detailed later. \par

In the FE layer, the displacement charge density $Q_\text{FE}$ can be written as 
\begin{equation} \label{eq:Qfe}
Q_\text{FE} = \epsilon_0\kappa E_\text{FE} + P,
\end{equation}
where $\epsilon_0$ is the vacuum dielectric constant, $\kappa$ is the background dielectric constant of the FE material~\cite{Tagantsev2008}, $E_\text{FE}$ is the electric field across the FE oxide, and $P$ is the average FE polarization. The dynamics of $P$ is governed by the Landau-Khalatnikov (LK) equation,
\begin{equation} \label{eq:LK}
\pdv{P}{t} = -L\qty( 2\alpha_1 P + 4\alpha_{11}P^3 + 6\alpha_{111}P^5 - E_\text{FE} ),
\end{equation}
where $\{\alpha_1, \alpha_{11}, \alpha_{111} \}$ are thermodynamic expansion coefficients for the bulk FE free energy, and $L$ is the kinetic coefficient, which is inversely proportional to domain viscosity~\cite{Landau1937,Ginzburg1945,Devonshire1949,Hsu2018,Chang2018}. Note that in this work, the free charge density $Q_\text{f}$ is equal to the FE displacement charge density $Q_\text{FE}$ with the assumption that the leakage of an FE thin film is negligible~\cite{Hoffmann2016}.

It was experimentally reported that the trapped charges may exist at the FE-DE interface through the DE leakage, which plays an important role in the electrical properties and DE breakdown of MFIM capacitors~\cite{Si2019}. In this regard, we introduce DE leakage and interfacial trapped charges into the MFM and MFIM systems. For simplicity, the DE leakage for $\abs{V_\text{DE}} > V_\text{c}$ is approximated using field-assisted tunneling mechanism according to the experiments on ultrathin \ch{Al2O3} (AO) dielectrics~\cite{Lin2005}, where $V_\text{DE}$ is the voltage across the DE layer and $V_\text{c}$ is the cutoff voltage below which the leakage current is negligible. $V_\text{c}$ is set to $\SI{0.2}{V}$ in this work. The leakage current when $\abs{V_\text{DE}}$ is above $V_\text{c}$ is therefore given by
\begin{equation} \label{eq:IL}
I_\text{L} = \text{sgn}\qty(V_\text{DE})\alpha E_\text{DE}^2\exp\qty(-\beta/\abs{E_\text{DE}})\cdot A,
\end{equation}
where $\alpha$ and $\beta$ are physical parameters related to the system band structures, $E_\text{DE}$ is the electric field across the DE oxide, and a sign function is used for the current direction.

To extract the DE charge density $\qty(Q_\text{DE})$ accumulated at the DE-metal interface, we integrate the displacement current through the DE layer with respect to time:
\begin{equation} \label{eq:Qde}
Q_\text{DE}\qty(t) = \frac{1}{A}\int_0^t I_\text{DE}\qty(\tau)\dd{\tau} + Q_\text{DE}\qty(0),
\end{equation}
where $Q_\text{DE}\qty(0)$ is the DE charges before switching. Note that $Q_\text{DE}\qty(0)$ is an initial condition in this framework and may vary from sample to sample depending on the experimental conditions. 

\begin{figure}[!t]
\centering
\includegraphics[width=0.45\textwidth]{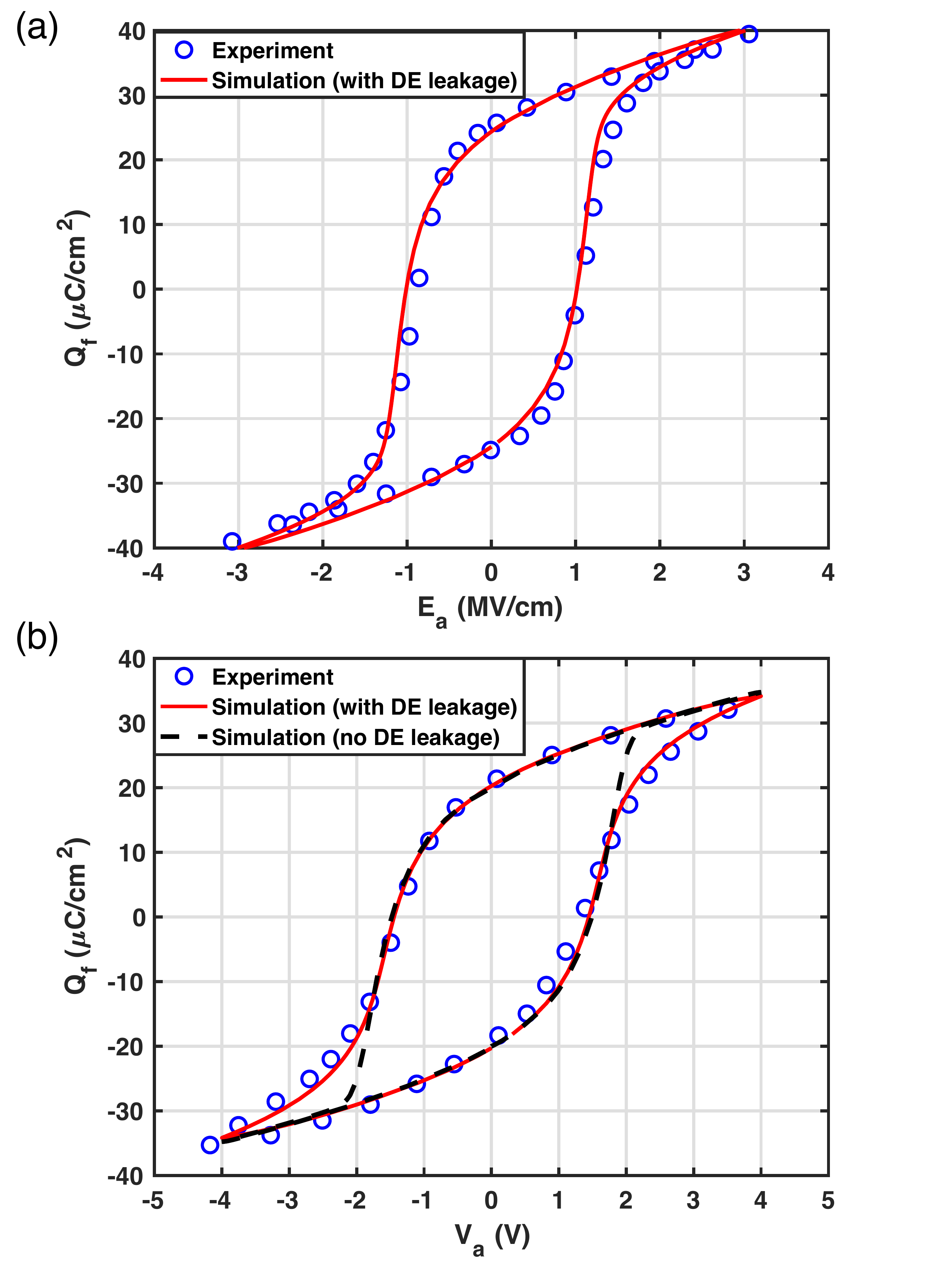}
\caption{(a) The simulated hysteresis loop of a TiN/11.3 nm HZO/TiN MFM capacitor in comparison with the experimental data extracted from Ref.~\cite{Hoffmann2018}. (b) The simulated hysteresis loop of a TiN/10 nm HZO/TiN MFM capacitor in comparison with the experimental data extracted from Ref.~\cite{Ni2018a}. The simulation without an interfacial oxide layer and the associated DE leakage is plotted for comparison.}
\label{fig:mfm}
\end{figure}

At the FD-DE interface, the charge neutrality condition has to be satisfied due to the displacement field continuity in the normal direction. Therefore, the interfacial trapped charge $Q_\text{it}$ can be obtained with the expression given by
\begin{equation} \label{eq:neutrality}
Q_\text{FE} +\widetilde{Q}_\text{DE} + Q_\text{it} = 0,
\end{equation}
where $\widetilde{Q}_\text{DE} = -Q_\text{DE}$ is the DE charge at the FE-DE interface.

\section{\label{sec:level1}Results and Discussion} \label{sec3}
Based on the physical model described in Sec.~\ref{sec2}, we study the transient charge responses of MFM and MFIM capacitors, including hysteresis loops and pulse switching dynamics. The underlying physics behind the recent experimental observations is discussed in detail. For convenience, the theoretical model is implemented in a circuit-compatible manner so that all the dynamic simulations can be performed accurately and self-consistently in the SPICE simulator. The detailed implementation can be found in~\cite{Hsu2020}.

\begin{table}[!t]  
\renewcommand{\arraystretch}{1.3}  
\centering
\caption{Parameters Used For MFM from Ref.~\cite{Hoffmann2018}} \label{tab:1}
\begin{threeparttable}
\begin{tabular}{c c}
\hline \hline
Parameter & Value\\ \hline
Area $\qty(A)$ \si{(\mu m^2)} & $\num{7000}$~\cite{Hoffmann2018} \\ \hline
FE thickness $\qty(t_\text{FE})$ \si{(nm)} & $\num{11.3}$~\cite{Hoffmann2018} \\ \hline
$\alpha_1$ (m/F) & \num{-1.43e8} \\ \hline
$\alpha_{11}$ \si{(m^5 C^{-2}/F)}  & \num{1.55e9} \\ \hline
$\alpha_{111}$ \si{(m^9 C^{-4}/F)} & \num{1.1e9} \\ \hline
$\kappa$ & 10 \\ \hline
$L$ $\qty(\qty[\si{\Omega m}]^{-1})$ & \num{3e-3} \\ \hline 
EOT $\qty(t_\text{DE})$ \si{(nm)} & 0.1 \\ \hline
$\alpha$ \si{(A/V^2)} & \num{9.57e-16} \\ \hline 
Frequency \si{(kHz)} & 10~\cite{Hoffmann2018} \\ \hline
Pulse amplitude \si{(V)} & 3.5~\cite{Hoffmann2018} \\ \hline \hline
\end{tabular}
\end{threeparttable}
\end{table}

\begin{figure*}[!t]
\centering
\includegraphics[width=1\textwidth]{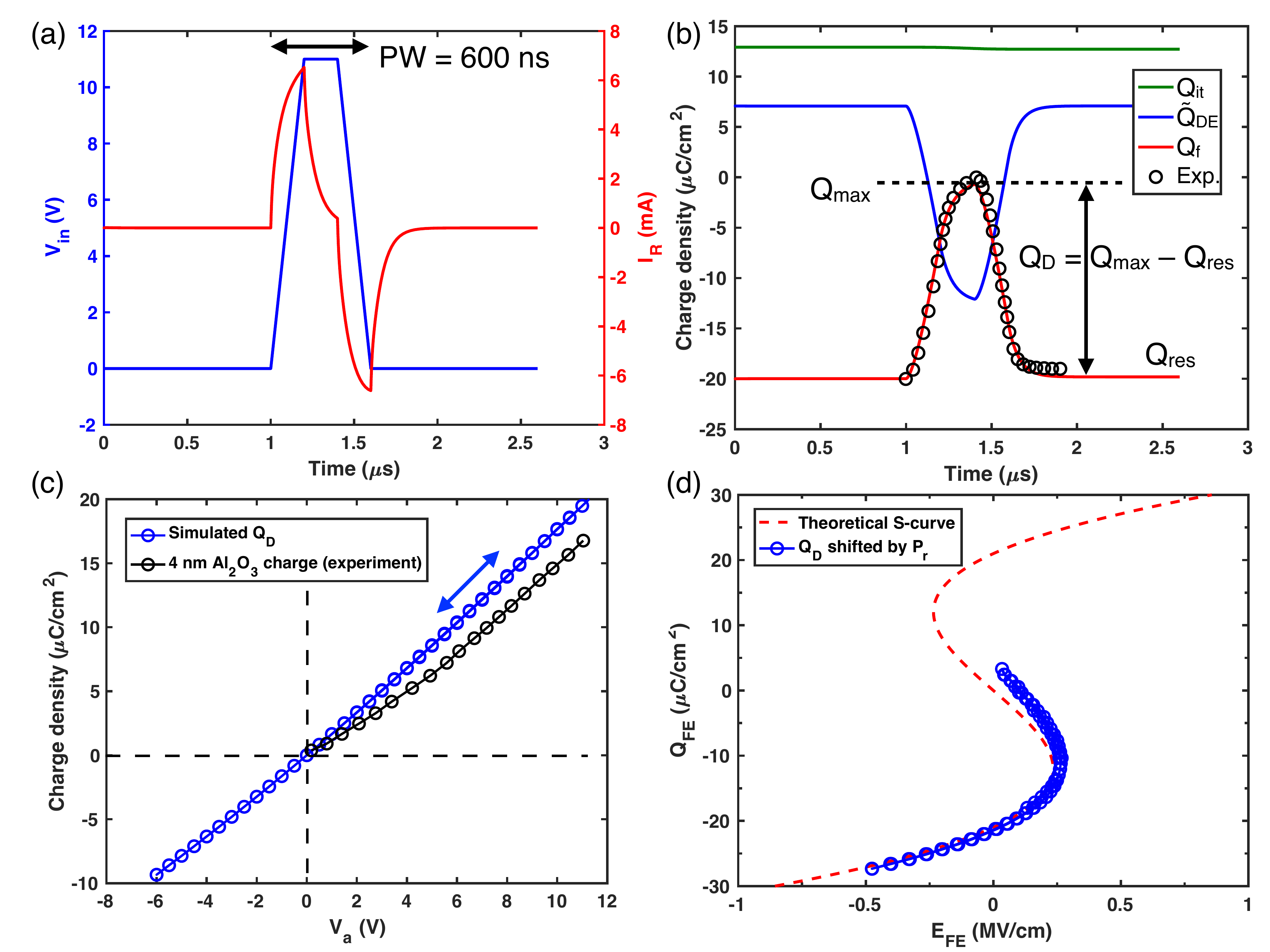}
\caption{(a) The applied pulse $V_\text{in}$ from Ref~\cite{Hoffmann2018} (left) and the simulated current flowing through the series resistor $I_\text{R}$ (right). (b) Simulated transient responses of the free charge $Q_\text{f}$ and the calculated DE charge $Q_\text{DE}$ and trapped charge $Q_\text{it}$ in comparison with the measured free charge response extracted from Ref.~\cite{Hoffmann2018}. The definitions of $Q_\text{max}$, $Q_\text{res}$ and $Q_\text{D}$ in the main text are also shown. (c) The release charge $Q_\text{D}$ at ascending and descending pulse trains with the amplitude $V_\text{a}$:  $\SI{0}{V} \to \SI{12}{V} \to \SI{-6}{V} \to \SI{0}{V}$, showing the hysteresis-free charge boost compared to the charges on the associated AO capacitor extracted from Ref.~\cite{Hoffmann2018}. (d) The simulated hysteresis-free S-curve compared to the static curve predicted by the Landau formalism using the FE material parameters obtained in Sec.~\ref{sec3a}.}
\label{fig:3}
\end{figure*}

\subsection{\label{sec:level2}MFM Capacitors} \label{sec3a}
The MFIM stack structure of interest is TiN/\ch{Hf_{0.5}Zr_{0.5}O2}/\ch{Al2O3}/TiN capacitors, where Zr-doped hafnium oxides (HZO) are a widely used FE material in the recent experimental measurements~\cite{Kobayashi2016,Ni2018a,Hoffmann2018,Hoffmann2019,Kim2019}. Before we study the MFIM capacitors, we first extract the FE parameters by simulating the hysteresis loops of \ch{TiN}/HZO/\ch{TiN} MFM capacitors based on the conventional polarization-voltage (P-V) measurements. As suggested by the experiments~\cite{Pesic2016,Kim2016a,Chouprik2019,Goh2020}, an interfacial oxide layer is likely to form between the FE layer and the metal contacts during the fabrication process of MFM capacitors. In this regard, an ultrathin DE layer is introduced at the FE-metal interface of an MFM capacitor. Note that the depolarization, which is caused by the finite screening effect of the metal contacts, manifests itself with the interfacial oxide layer~\cite{Lomenzo2020}. Based on the P-V measurement in Ref.~\cite{Hoffmann2018}, the equivalent oxide thickness (EOT) of the interfacial oxide layer is found to be around $\SI{0.1}{nm}$ given that the dielectric constant of \ch{SiO2} ($\epsilon_\text{\ch{SiO2}}$) is $3.9$. The corresponding interfacial capacitance $C_\text{DE}=\epsilon_0\epsilon_\text{\ch{SiO2}}/\text{EOT} \approx \SI{0.35}{F/m^2}$, which is consistent with the reported value for the FE/\ch{TiN} interface~\cite{Lomenzo2020}. The physical thickness of the interfacial oxide layer can be estimated to be \SI{0.9}{nm} given the relative permittivity of $36$ for the tetragonal and orthorhombic phase in \ch{HfO2} and \ch{ZrO2}~\cite{Lomenzo2020}. Fig.~\ref{fig:mfm}(a) shows the simulated hysteresis loop of an MFM capacitor based on the experimental P-V measurement in Ref.~\cite{Hoffmann2018}. To further confirm the effect of an oxide layer between the FE and the metal contact, we simulate the hysteresis loop of an MFM capacitor from another measurement ~\cite{Ni2018a}, as shown in Fig.~\ref{fig:mfm}(b). The simulation without considering the interfacial oxide layer is also plotted for comparison. Our simulations indicate that the oxide layer at the FE-metal interface may be one of the important factors that could influence the hysteresis shape of MFM. The parameters extracted based on the experimental P-V measurement in Ref.~\cite{Hoffmann2018} are summarized in Table~\ref{tab:1}. The FE properties of \SI{11.3}{nm} HZO are used for the study of MFIM capacitors. \par

\begin{figure}[!t]
\centering
\includegraphics[width=0.45\textwidth]{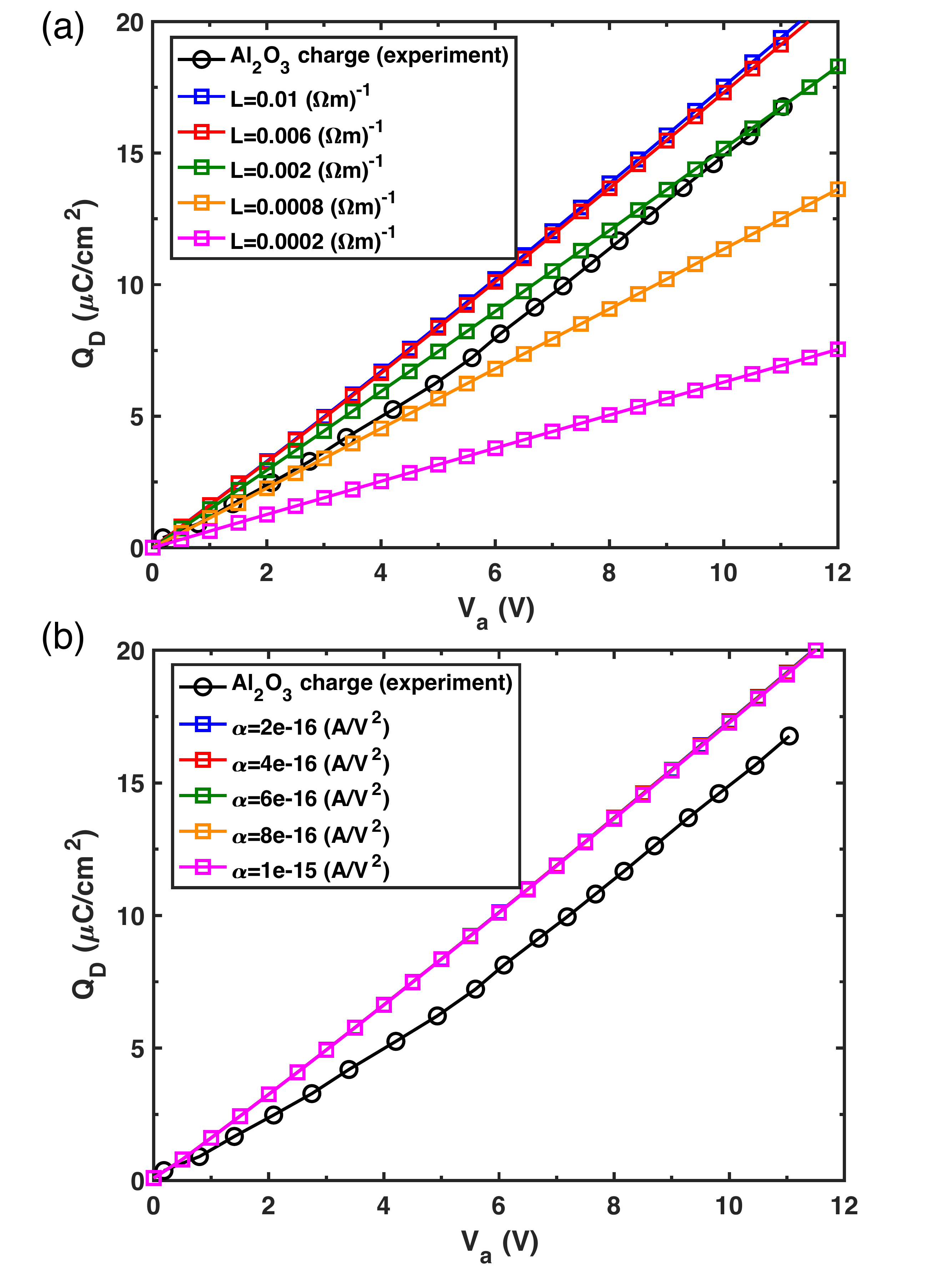}
\caption{(a) The effect of $L$ on the charge boost effect. As $L$ increases, the release charge $Q_\text{D}$ on MFIM gradually becomes larger than that on the associated DE capacitor. (b) The effect of $\alpha$ on the charge boost, which indicates that the charge boost is not dominated by the DE leakage. The charges on AO are extracted from the measurements in Ref.~\cite{Hoffmann2018}.}
\label{fig:4}
\end{figure}

\begin{figure}[!t]
\centering
\includegraphics[width=0.45\textwidth]{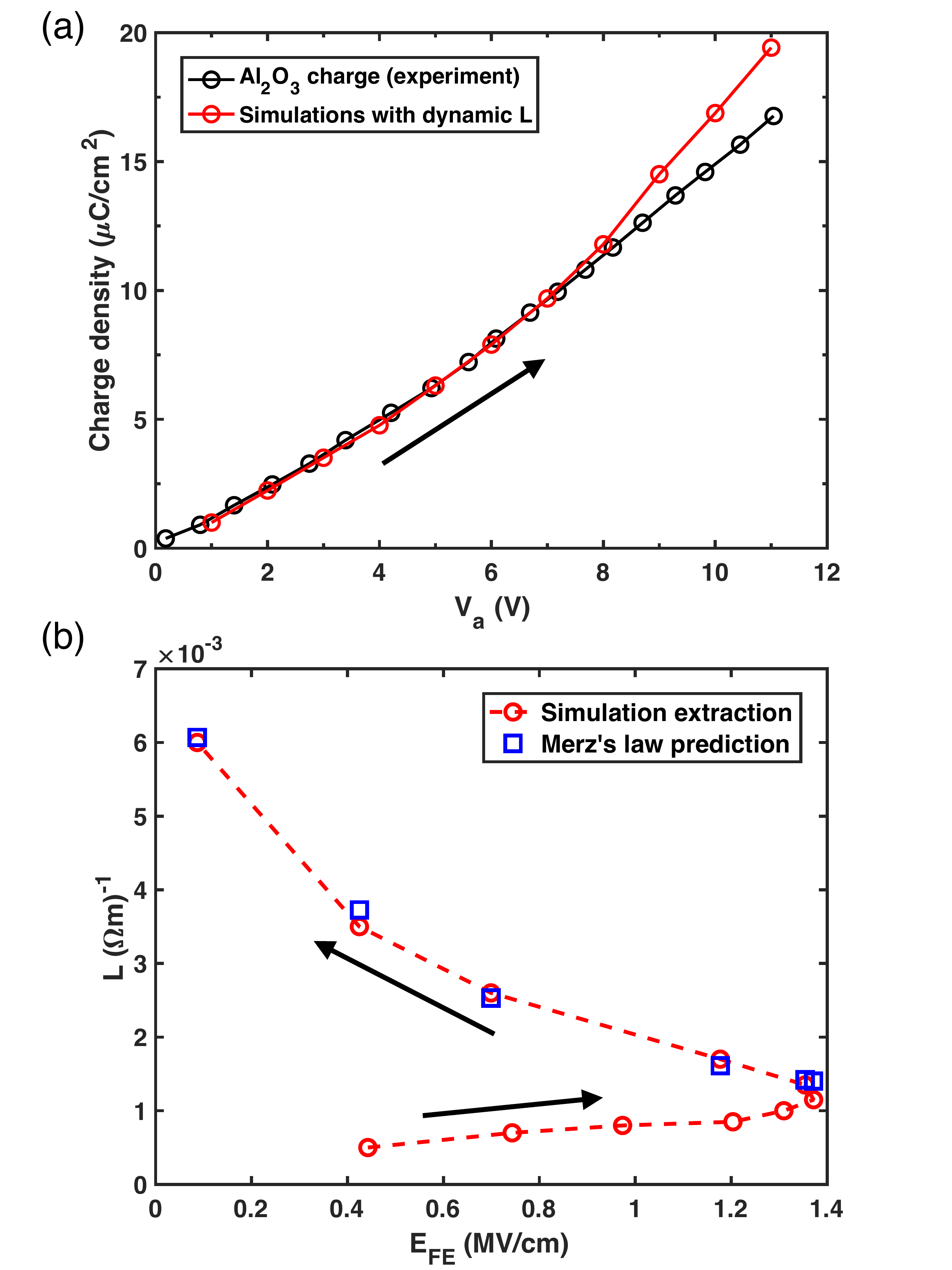}
\caption{(a) The corresponding $Q_\text{D}$ using $L$ extracted based on the charge responses in Ref.~\cite{Hoffmann2018}, where charges on AO are from Ref.~\cite{Hoffmann2018} as well. (b) The extracted $L$ in the NC region compared to the Merz's law prediction with $E_0 = \SI{0.12}{MV/cm}$. The turning point of $E_\text{FE}$ indicates the onset of NC region. The black arrows in (a) and (b) indicate the applied voltage sweep direction.}
\label{fig:L_ext}
\end{figure}

\begin{figure*}[!t]
\centering
\includegraphics[width=1\textwidth]{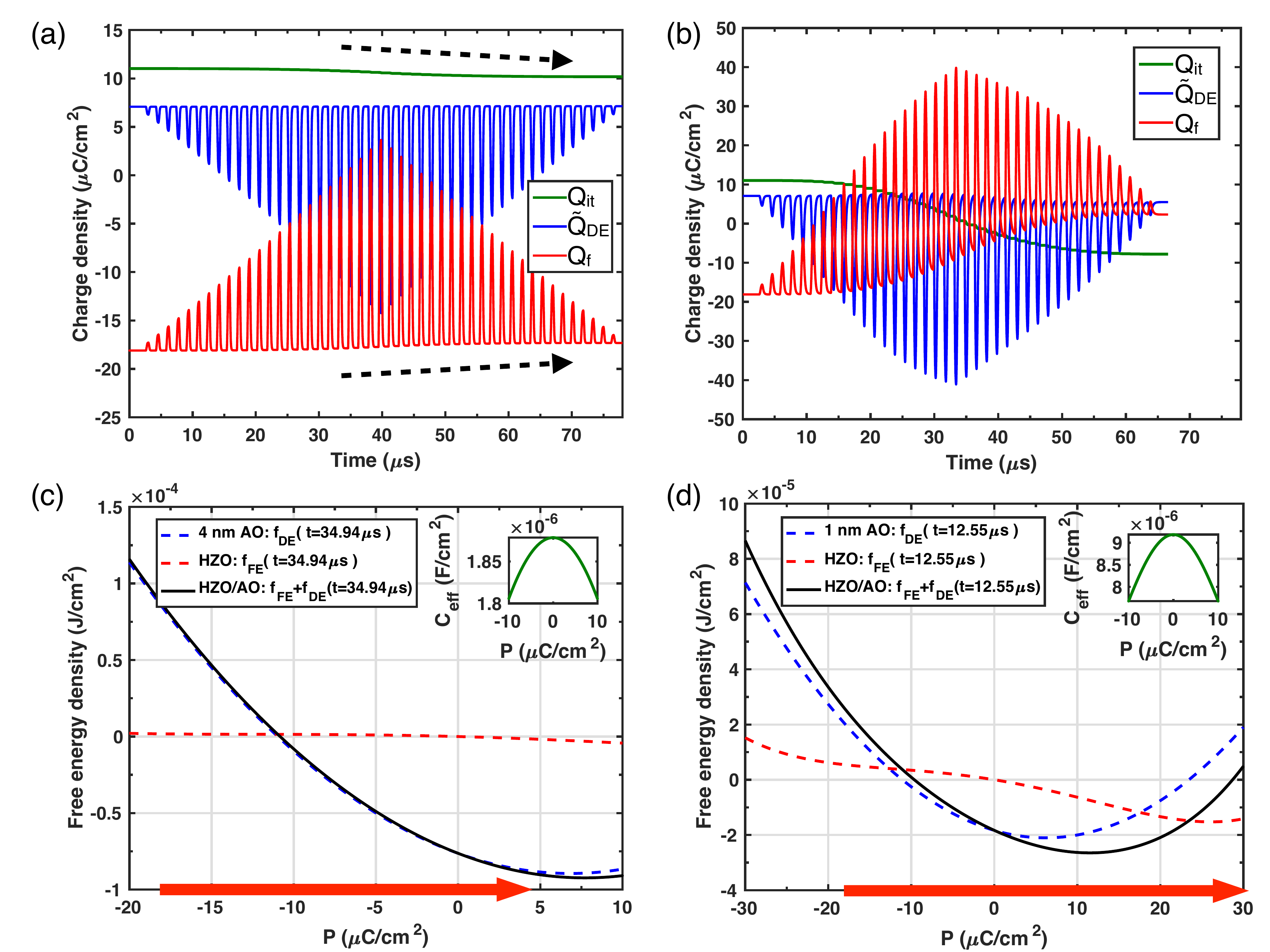}
\caption{(a) Charge responses of the \SI{11.3}{nm} HZO/\SI{4}{nm} AO stack at ascending and descending pulse trains with the amplitude $V_\text{a}$: $\SI{0}{V} \to \SI{12}{V} \to \SI{0}{V}$. The dashed arrows indicate the trapped charge leak and the increasing residual charges during the pulse trains. $\alpha = \SI{2e-16}{A/V^2}$ for the \SI{4}{nm} AO stack. (b) Charge responses of the \SI{11.3}{nm} HZO/\SI{4}{nm} AO stack at ascending and descending pulse trains with $V_\text{a}$: $\SI{0}{V} \to \SI{10}{V} \to \SI{0}{V}$. $\alpha = \SI{7.57e-16}{A/V^2}$ for the \SI{1}{nm} AO stack. (c) The free energy profiles of the \SI{11.3}{nm} HZO, the \SI{4}{nm} AO, and the corresponding HZO/AO stack near zero polarization at $t=t_0$. The red arrow indicates the switching range of $P$ during the simulation. (d) The free energy profiles of the \SI{11.3}{nm} HZO, the \SI{1}{nm} AO, and the corresponding HZO/AO stack near zero polarization at $t=t_0$. The red arrow indicates the switching range of $P$ during the simulation.}
\label{fig:hysteresis}
\end{figure*}

\subsection{\label{sec:level2}MFIM Capacitors}
To study the TiN/HZO/AO/TiN capacitor, we apply the parameters obtained from the preceding section for \SI{11.3}{nm} HZO. Based on the experimental setup in Ref.~\cite{Hoffmann2018}, ascending and descending pulse trains with a \SI{600}{ns} pulse width (PW) were applied to the MFIM capacitor with \SI{11.3}{nm} HZO and \SI{4}{nm} AO. One of the voltage pulses and the total current flowing through the external resistor are shown in Fig.~\ref{fig:3}(a). Fig.~\ref{fig:3}(b) shows the corresponding transient free charge responses from the simulation. In the steady state before the pulse voltage is applied, the FE layer is in the negative remanent polarization $(P_\text{r})$ state, as suggested in Ref.~\cite{Hoffmann2018}. To compensate the FE charge density, there must be the same amount of charges with opposite signs at the FE-DE interface. However, a dielectric material like AO cannot support a charge density as large as around $\SI{20}{\mu C/cm^2}$ for such a long period of time without breakdown~\cite{Lin2005,Si2019}. Therefore, here the initial DE charge $\widetilde{Q}_\text{DE}(0)$ is assumed to be $\SI{7}{\mu C/cm^2}$ reported in Ref.~\cite{Si2019}, and the rest of the compensating charges are from the interfacial trapped charge $Q_\text{it}$~\cite{Si2019}. Such a large amount of trapped charges are likely to be introduced through the DE leakage to the FE-DE interface during the wake-up process after the fabrication. At a large pulse amplitude, a large amount of negative DE charges accumulated at the FE-DE interface indicate that most of the applied voltage is across the DE layer in this stack system. Thus, in such pulse measurements, the FE polarization is barely switched even if a large voltage is applied. Similar to the experiments, the release charge $Q_D$ is defined as the difference between the maximum charge $Q_\text{max}$ and the residual charge $Q_\text{res}$ at a given pulse amplitude $V_\text{a}$; that is, $Q_D = Q_\text{max} - Q_\text{res}$. At a fixed kinetic coefficient $L=\SI{6e-3}{(\Omega m)^{-1}}$, it is found that $Q_D$ of MFIM is enhanced without hysteresis compared to the measured charges on the associated standalone \SI{4}{nm} AO capacitor, as shown in Fig.~\ref{fig:3}(c). From the simulation, the electric field across the FE oxide, $E_\text{FE} = V_\text{FE}/t_\text{FE}$, can be directly obtained, and the theoretical static curve is derived from Eq.~\eqref{eq:LK} in the steady state:
\begin{equation} \label{eq:s-curve}
E_\text{FE} = 2\alpha_1 P + 4\alpha_{11}P^3 + 6\alpha_{111}P^5.
\end{equation}
Fig.~\ref{fig:3}(d) shows that an S-curve with negligible hysteresis can be achieved with the DE leakage parameter $\alpha = \SI{2e-16}{A/V^2}$. For the static curve, $Q_D$ is shifted by $P_\text{r}$, which is the initial reference point and was reported as around \SI{18}{\mu C/cm^2}~\cite{Hoffmann2018}. Note that the S-curve obtained from the simulation does not go through the origin because $L$ for the FE response has not been calibrated dynamically based on the polarization charge responses. \par

In Fig.~\ref{fig:4}(a), we show that the experimentally observed charge boost is mainly determined by the kinetic coefficient $L$. The increasing $L$ indicates the faster FE charge responses to the applied voltage, leading to larger maximum charges $Q_\text{max}$ in the same PW. Hence, the release charge $Q_\text{D}$ can be enhanced with less FE domain viscosity. To further confirm this finding, we also explore the effect of DE leakage $\alpha$ and find that the DE leakage is not a dominant factor in the observed charge boost effect, as shown in Fig.~\ref{fig:4}(b). Note that for simplicity, $\beta$ in Eq.~\eqref{eq:IL} is kept fixed at $\SI{3.7e-7}{V/m}$ for all the simulations in this work due to the much less dominant effect in the systems. \par
Based on the experimentally measured free charge response from Ref.~\cite{Hoffmann2018}, the kinetic coefficient $L$ can be numerically extracted at a given pulse amplitude. With extracted $L$, Fig.~\ref{fig:L_ext}(a) demonstrates that the charge boost is enhanced when the applied pulse is large enough to induce the FE response, whereas a small applied pulse only results in the DE response in the FE layer and thus no charge boost is observed. As a reciprocal of the domain viscosity, $L$ is assumed to be proportional to the domain wall mobility $\mu_\text{d}$ corresponding to the domain nucleation; that is, $L \propto \mu_\text{d}$. Based on Merz's law~\cite{Merz1956}, the relationship between $L$ and nonzero $E_\text{FE}$ can be derived: 
\begin{equation} \label{eq:L_Merz}
L \propto \frac{\mu_\text{d}\abs{E_\text{FE}}}{\abs{E_\text{FE}}} \propto \frac{1}{\abs{E_\text{FE}}}v_\text{d} \propto \frac{1}{\abs{E_\text{FE}}} \frac{1}{t_\text{s}} \propto \frac{1}{\abs{E_\text{FE}}} e^{-E_0/\abs{E_\text{FE}}},
\end{equation}
where $v_\text{d}$ is the domain wall velocity inversely proportional to the switching characteristic time $t_\text{s}$, and $E_0$ is the activation field for domain switching~\cite{Merz1956}. Fig.~\ref{fig:L_ext}(b) shows that when the FE is driven into the NC region, $L$ can be described by Merz's law, which suggests the domains are likely to be in the creep region as the electric field across the FE $\qty(E_\text{FE})$ becomes smaller~\cite{Jo2009}. From Eq.~\eqref{eq:L_Merz}, as $\abs{E_\text{FE}}$ approaches $0^+$, $L$ is asymptotic to zero because of zero driving force for domain switching. Before FE domain switching, only the DE response is induced, and $L$ is found to be linearly dependent on $E_\text{FE}$. \par

In contrast to the TiN/\SI{11.3}{nm} HZO/\SI{4}{nm} AO/TiN capacitor, the charge hysteresis was observed in a TiN/\SI{11.3}{nm} HZO/\SI{1}{nm} AO/TiN capacitor~\cite{Hoffmann2018}. It was suggested that the appearance of such hysteresis was attributed to the capacitance mismatch between the FE and the DE layer as the DE thickness decreases based on the theory of NC stabilization. However, here we demonstrate that such hysteresis is caused by the DE leakage and the associated interfacial trapped charge dynamics instead of NC stabilization. At the pulse trains with ascending and descending amplitudes, Fig.~\ref{fig:hysteresis}(a) and (b) show the charge responses of the \SI{11.3}{nm} HZO/\SI{4}{nm} AO stack and the \SI{11.3}{nm} HZO/\SI{1}{nm} AO stack, respectively. For the \SI{11.3}{nm} HZO/\SI{4}{nm} AO stack, when a large voltage is across the DE layer, DE leakage is induced and therefore, the trapped charges leak out through the DE, as can be seen from the slightly decreased $Q_\text{it}$ in Fig.~\ref{fig:hysteresis}(a). As a result, the residual charges of $Q_\text{f}$ after each pulse gradually increase in Fig.~\ref{fig:hysteresis}(a). For the \SI{11.3}{nm} HZO/\SI{1}{nm} AO stack in Fig.~\ref{fig:hysteresis}(b), the leakage effect becomes more evident compared to the \SI{11.3}{nm} HZO/\SI{4}{nm} AO stack. The simulations indicate that the possible cause of the increasing residual charges after pulse switching experimentally observed in Ref.~\cite{Hoffmann2018}, especially at a large pulse amplitude, can be explained by the transient responses of trapped charges at the FE-DE interface. Furthermore, from Fig.~\ref{fig:hysteresis}(a) and (b) and Eq.~\eqref{eq:Qde}, one can infer that the role of the initial DE charge $\widetilde{Q}_\text{DE}\qty(0)$ plays in the transient responses is a shift in $\widetilde{Q}_\text{DE}$ and $Q_\text{it}$ dynamics, which means that the choice of the $\widetilde{Q}_\text{DE}\qty(0)$ value does not affect the conclusions drawn from the simulations. \par

The hysteresis behavior can be studied with the curvature of the total free energy profile of the system near polarization reversal. The Landau free energy of the FE in a unit of \si{J/m^2} is given by
\begin{equation} \label{eq:f_FE}
f_\text{FE} = \qty(\alpha_1 P^2 + \alpha_{11} P^4 + \alpha_{111} P^6 - E_\text{FE} P) t_\text{FE}.
\end{equation}
The energy stored in a capacitor in terms of $P$ can be derived as 
\begin{equation} \label{eq:f_DE}
\begin{split}
f_\text{DE} &= Q_\text{DE}^2/(2C_\text{DE}) - Q_\text{DE}V_\text{DE} \\
&= \qty(P - \Delta Q)^2/(2C_\text{DE}) - \qty(P - \Delta Q)V_\text{DE}, 
\end{split}
\end{equation}
where $C_\text{DE}$ is the DE capacitance per unit area and $\Delta{Q} = P - Q_\text{DE}$ is the difference between polarization and the DE charges, which is caused by the interfacial trapped charges. Fig.~\ref{fig:hysteresis}(c) and (d) show the free energy profiles of the FE, the DE, and the FE-DE stacks at $t=t_0$, where $t_0$ is the time at which $P$ is around $\SI{0}{\mu C/cm^2}$. Note that in the presence of the trapped charges, the FE and DE energy profiles do not align at $P=\SI{0}{\mu C/cm^2}$ because the FE charges are not totally compensated by the DE charges. In the \SI{11.3}{nm} HZO/\SI{4}{nm} AO stack, polarization $P$ is switched between one of the energy minimums and the energy maximum, as indicated by the red arrow in Fig.~\ref{fig:hysteresis}(c). In other words, $P$ does not go to the other energy minimum state. Note that when $P$ is around the energy maximum, the FE is driven into its NC region, where the curvature of the free energy profile (and the FE capacitance) is negative. The effective capacitance $C_\text{eff}$ of the MFIM system can be obtained by calculating the curvature of the total free energy profile:
\begin{equation} \label{eq:c_eff}
C_\text{eff} = \qty(\pdv[2]{f_\text{DE}}{P} + \pdv[2]{f_\text{FE}}{P})^{-1}.
\end{equation}
From Eq.~\eqref{eq:c_eff}, $C_\text{eff}$ is positive when the positive DE capacitance is dominant over the negative FE capacitance. With a positive $C_\text{eff}$, the NC region gets stabilized and no hysteresis is expected according to the NC theory~\cite{Salahuddin2008,Khan2016}. Note that $\Delta Q$ in Eq.~\eqref{eq:f_DE} (and thus trapped charges) does not affect the effective capacitance. \par

\begin{figure}[!t]
\centering
\includegraphics[width=0.45\textwidth]{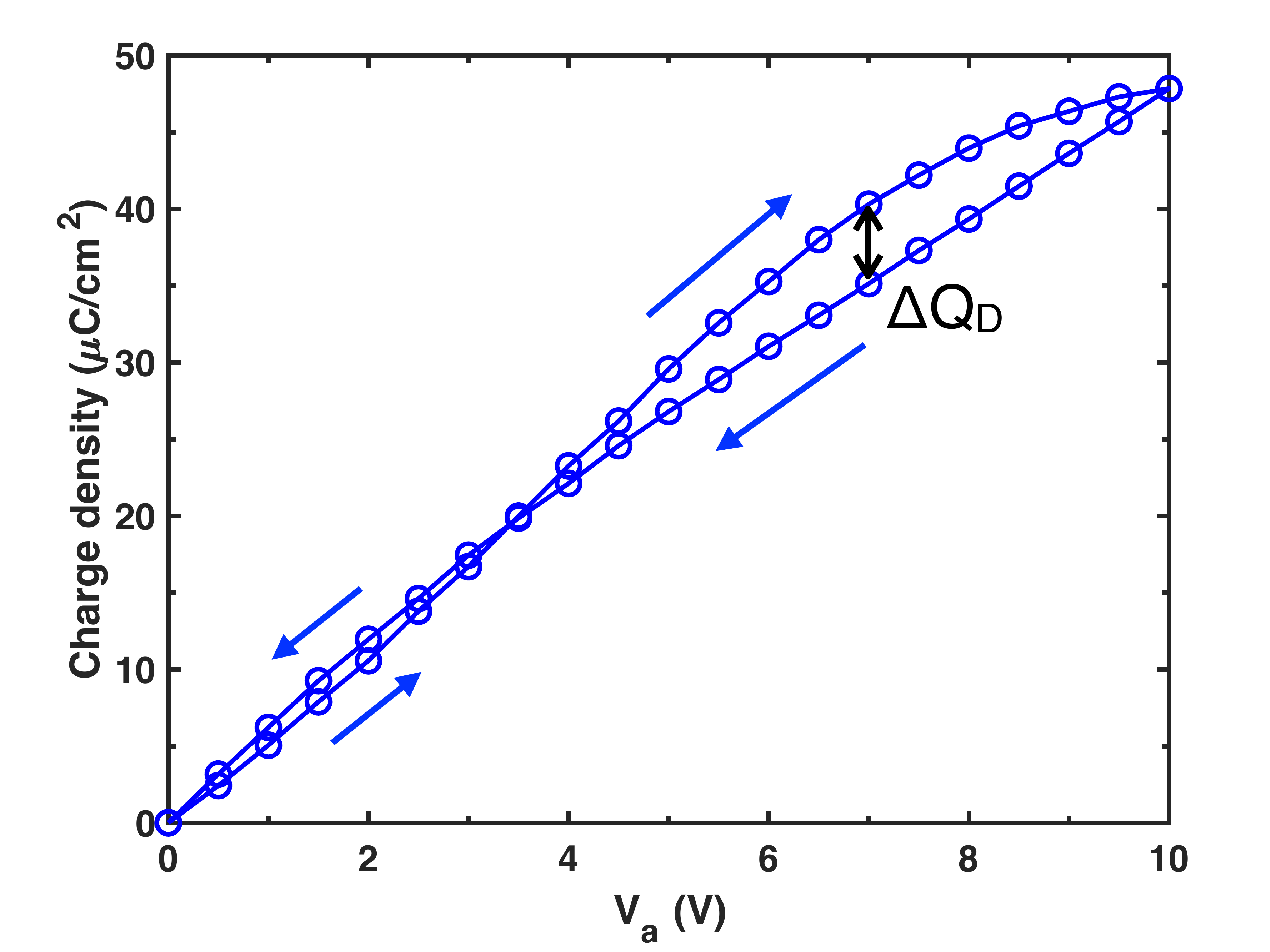}
\caption{The release charge $Q_\text{D}$ of the \SI{11.3}{nm} HZO/\SI{1}{nm} AO stack with $\alpha = \SI{7.57e-16}{A/V^2}$ shows clear hysteresis in a clockwise manner, as observed in Ref.~\cite{Hoffmann2018}.}
\label{fig:hysteresis_1nm}
\end{figure}

As shown in the inset of Fig.~\ref{fig:hysteresis}(c), the effective capacitance of the \SI{4}{nm} AO stack is positive around zero polarization, indicating the FE-DE capacitance match and thus hysteresis-free NC stabilization. For the \SI{1}{nm} AO stack, the effective capacitance is also positive around zero polarization, as shown in the inset of Fig.~\ref{fig:hysteresis}(d). Therefore, based on the theory of NC stabilization, such a \SI{1}{nm} AO stack is not expected to show hysteresis. However, consistent with the experimental findings in Ref.~\cite{Hoffmann2018}, the release charge $Q_\text{D}$ of such a \SI{1}{nm} AO stack shows clear hysteresis with the DE leakage parameter $\alpha = \SI{7.57e-16}{A/V^2}$ in the simulation, as shown in Fig.~\ref{fig:hysteresis_1nm}. Note that $\alpha$ of the \SI{1}{nm} AO stack is larger than that of the \SI{4}{nm} AO stack due to the larger leakage current in a thinner DE. In contrast to the \SI{4}{nm} AO stack, the FE polarization of the \SI{1}{nm} AO stack is driven into the other energy minimum state in the measurement, as indicated by the red arrow in Fig.~\ref{fig:hysteresis}(d). The appearance of hysteresis in the \SI{1}{nm} AO stack can be attributed to the increasing FE charge responses due to the larger DE leakage of the thinner DE compared to the \SI{4}{nm} AO stack. Therefore, we show that the hysteresis observed in an MFIM stack with a reduced DE thickness is caused not by the FE-DE capacitance mismatch but by the fact that the FE polarization transitions from one state to the other state with the help of DE leakage. This finding is consistent with the previous experimental observation of FE-gated Ge p-channel transistors~\cite{Zhou2019}. In addition, Fig.~\ref{fig:hysteresis_1nm} shows that, for the MFIM with \SI{1}{nm} AO, the maximum release charge difference in the hysteresis $\Delta Q_\text{D} \approx \SI{5.2}{\mu C/cm^2}$ at the pulse amplitude of \SI{7}{V}, which is close to the reported value of \SI{4.7}{\mu C/cm^2} in Ref.~\cite{Hoffmann2018}. Note that the hysteresis direction is in a clockwise manner, which is also consistent with the experimental measurements in Ref.~\cite{Hoffmann2018}.

\vspace{-0.5cm}
\section{\label{sec:level1}Conclusion} \label{sec4}
In summary, this paper presents underlying physical mechanisms for the hysteresis-free negative capacitance effect experimentally observed in MFIM capacitors. With the DE leakage and interfacial trapped charges included, we extract the FE parameters by capturing the measured hysteresis loops of MFM capacitors. More importantly, for MFIM stacks, we show (i) the charge boost is mainly determined by the faster FE domain responses near the polarization reversal and (ii) the charge responses would show hysteresis as long as the FE polarization is switched from one state to the other with the aid of DE leakage. The analyses of thermodynamic energy profiles in the presence of interfacial trapped charges indicate that the observed hysteresis in MFIM is caused by the FE polarization switching rather than the FE-DE capacitance mismatch suggested by the theory of NC stabilization. The theoretical demonstrations provide important physical insights into the alleged static NC effect for emerging low-power logic devices.

\vspace{0.3cm}
\section*{\label{sec:level1}Acknowledgments}
This work was funded by Intel Corporation through Semiconductor Research Corporation MSR-INTEL TASK 2835.001.

\RaggedRight
\bibliography{./MFIM}

\begin{thebibliography}{38}%
\makeatletter
\providecommand \@ifxundefined [1]{%
 \@ifx{#1\undefined}
}%
\providecommand \@ifnum [1]{%
 \ifnum #1\expandafter \@firstoftwo
 \else \expandafter \@secondoftwo
 \fi
}%
\providecommand \@ifx [1]{%
 \ifx #1\expandafter \@firstoftwo
 \else \expandafter \@secondoftwo
 \fi
}%
\providecommand \natexlab [1]{#1}%
\providecommand \enquote  [1]{``#1''}%
\providecommand \bibnamefont  [1]{#1}%
\providecommand \bibfnamefont [1]{#1}%
\providecommand \citenamefont [1]{#1}%
\providecommand \href@noop [0]{\@secondoftwo}%
\providecommand \href [0]{\begingroup \@sanitize@url \@href}%
\providecommand \@href[1]{\@@startlink{#1}\@@href}%
\providecommand \@@href[1]{\endgroup#1\@@endlink}%
\providecommand \@sanitize@url [0]{\catcode `\\12\catcode `\$12\catcode
  `\&12\catcode `\#12\catcode `\^12\catcode `\_12\catcode `\%12\relax}%
\providecommand \@@startlink[1]{}%
\providecommand \@@endlink[0]{}%
\providecommand \url  [0]{\begingroup\@sanitize@url \@url }%
\providecommand \@url [1]{\endgroup\@href {#1}{\urlprefix }}%
\providecommand \urlprefix  [0]{URL }%
\providecommand \Eprint [0]{\href }%
\providecommand \doibase [0]{http://dx.doi.org/}%
\providecommand \selectlanguage [0]{\@gobble}%
\providecommand \bibinfo  [0]{\@secondoftwo}%
\providecommand \bibfield  [0]{\@secondoftwo}%
\providecommand \translation [1]{[#1]}%
\providecommand \BibitemOpen [0]{}%
\providecommand \bibitemStop [0]{}%
\providecommand \bibitemNoStop [0]{.\EOS\space}%
\providecommand \EOS [0]{\spacefactor3000\relax}%
\providecommand \BibitemShut  [1]{\csname bibitem#1\endcsname}%
\let\auto@bib@innerbib\@empty
\bibitem [{\citenamefont {Moore}(1998)}]{Moore1998}%
  \BibitemOpen
  \bibfield  {author} {\bibinfo {author} {\bibfnamefont {G.}~\bibnamefont
  {Moore}},\ }\href {\doibase 10.1109/jproc.1998.658762} {\bibfield  {journal}
  {\bibinfo  {journal} {Proceedings of the {IEEE}}\ }\textbf {\bibinfo {volume}
  {86}},\ \bibinfo {pages} {82} (\bibinfo {year} {1998})},\ \bibinfo {note}
  {doi: 10.1109/jproc.1998.658762}\BibitemShut {NoStop}%
\bibitem [{\citenamefont {Kim}\ \emph {et~al.}(2003)\citenamefont {Kim},
  \citenamefont {Austin}, \citenamefont {Blaauw}, \citenamefont {Mudge},
  \citenamefont {Flautner}, \citenamefont {Hu}, \citenamefont {Irwin},
  \citenamefont {Kandemir},\ and\ \citenamefont {Narayanan}}]{Kim2003}%
  \BibitemOpen
  \bibfield  {author} {\bibinfo {author} {\bibfnamefont {N.~S.}\ \bibnamefont
  {Kim}}, \bibinfo {author} {\bibfnamefont {T.}~\bibnamefont {Austin}},
  \bibinfo {author} {\bibfnamefont {D.}~\bibnamefont {Blaauw}}, \bibinfo
  {author} {\bibfnamefont {T.}~\bibnamefont {Mudge}}, \bibinfo {author}
  {\bibfnamefont {K.}~\bibnamefont {Flautner}}, \bibinfo {author}
  {\bibfnamefont {J.~S.}\ \bibnamefont {Hu}}, \bibinfo {author} {\bibfnamefont
  {M.}~\bibnamefont {Irwin}}, \bibinfo {author} {\bibfnamefont
  {M.}~\bibnamefont {Kandemir}}, \ and\ \bibinfo {author} {\bibfnamefont
  {V.}~\bibnamefont {Narayanan}},\ }\href {\doibase 10.1109/mc.2003.1250885}
  {\bibfield  {journal} {\bibinfo  {journal} {Computer}\ }\textbf {\bibinfo
  {volume} {36}},\ \bibinfo {pages} {68} (\bibinfo {year} {2003})}\BibitemShut
  {NoStop}%
\bibitem [{\citenamefont {Salahuddin}\ and\ \citenamefont
  {Datta}(2008)}]{Salahuddin2008}%
  \BibitemOpen
  \bibfield  {author} {\bibinfo {author} {\bibfnamefont {S.}~\bibnamefont
  {Salahuddin}}\ and\ \bibinfo {author} {\bibfnamefont {S.}~\bibnamefont
  {Datta}},\ }\href@noop {} {\bibfield  {journal} {\bibinfo  {journal} {Nano
  Lett.}\ }\textbf {\bibinfo {volume} {8}},\ \bibinfo {pages} {405} (\bibinfo
  {year} {2008})},\ \bibinfo {note} {doi: 10.1021/nl071804g},\ \Eprint
  {http://arxiv.org/abs/http://dx.doi.org/10.1021/nl071804g}
  {http://dx.doi.org/10.1021/nl071804g} \BibitemShut {NoStop}%
\bibitem [{\citenamefont {Chang}\ \emph {et~al.}(2017)\citenamefont {Chang},
  \citenamefont {Avci}, \citenamefont {Nikonov},\ and\ \citenamefont
  {Young}}]{Chang2017a}%
  \BibitemOpen
  \bibfield  {author} {\bibinfo {author} {\bibfnamefont {S.-C.}\ \bibnamefont
  {Chang}}, \bibinfo {author} {\bibfnamefont {U.~E.}\ \bibnamefont {Avci}},
  \bibinfo {author} {\bibfnamefont {D.~E.}\ \bibnamefont {Nikonov}}, \ and\
  \bibinfo {author} {\bibfnamefont {I.~A.}\ \bibnamefont {Young}},\ }\href
  {\doibase 10.1109/jxcdc.2017.2750108} {\bibfield  {journal} {\bibinfo
  {journal} {{IEEE} Journal on Exploratory Solid-State Computational Devices
  and Circuits}\ }\textbf {\bibinfo {volume} {3}},\ \bibinfo {pages} {56}
  (\bibinfo {year} {2017})},\ \bibinfo {note} {doi:
  10.1109/jxcdc.2017.2750108}\BibitemShut {NoStop}%
\bibitem [{\citenamefont {Hsu}\ \emph {et~al.}(2018)\citenamefont {Hsu},
  \citenamefont {Pan},\ and\ \citenamefont {Naeemi}}]{Hsu2018}%
  \BibitemOpen
  \bibfield  {author} {\bibinfo {author} {\bibfnamefont {C.-S.}\ \bibnamefont
  {Hsu}}, \bibinfo {author} {\bibfnamefont {C.}~\bibnamefont {Pan}}, \ and\
  \bibinfo {author} {\bibfnamefont {A.}~\bibnamefont {Naeemi}},\ }\href
  {\doibase 10.1109/led.2018.2820118} {\bibfield  {journal} {\bibinfo
  {journal} {{IEEE} Electron Device Letters}\ }\textbf {\bibinfo {volume}
  {39}},\ \bibinfo {pages} {765} (\bibinfo {year} {2018})},\ \bibinfo {note}
  {doi: 10.1109/led.2018.2820118}\BibitemShut {NoStop}%
\bibitem [{\citenamefont {Ni}\ \emph {et~al.}(2018{\natexlab{a}})\citenamefont
  {Ni}, \citenamefont {Sharma}, \citenamefont {Zhang}, \citenamefont {Jerry},
  \citenamefont {Smith}, \citenamefont {Tapily}, \citenamefont {Clark},
  \citenamefont {Mahapatra},\ and\ \citenamefont {Datta}}]{Ni2018a}%
  \BibitemOpen
  \bibfield  {author} {\bibinfo {author} {\bibfnamefont {K.}~\bibnamefont
  {Ni}}, \bibinfo {author} {\bibfnamefont {P.}~\bibnamefont {Sharma}}, \bibinfo
  {author} {\bibfnamefont {J.}~\bibnamefont {Zhang}}, \bibinfo {author}
  {\bibfnamefont {M.}~\bibnamefont {Jerry}}, \bibinfo {author} {\bibfnamefont
  {J.~A.}\ \bibnamefont {Smith}}, \bibinfo {author} {\bibfnamefont
  {K.}~\bibnamefont {Tapily}}, \bibinfo {author} {\bibfnamefont
  {R.}~\bibnamefont {Clark}}, \bibinfo {author} {\bibfnamefont
  {S.}~\bibnamefont {Mahapatra}}, \ and\ \bibinfo {author} {\bibfnamefont
  {S.}~\bibnamefont {Datta}},\ }\href {\doibase 10.1109/ted.2018.2829122}
  {\bibfield  {journal} {\bibinfo  {journal} {{IEEE} Transactions on Electron
  Devices}\ }\textbf {\bibinfo {volume} {65}},\ \bibinfo {pages} {2461}
  (\bibinfo {year} {2018}{\natexlab{a}})}\BibitemShut {NoStop}%
\bibitem [{\citenamefont {Ni}\ \emph {et~al.}(2018{\natexlab{b}})\citenamefont
  {Ni}, \citenamefont {Jerry}, \citenamefont {Smith},\ and\ \citenamefont
  {Datta}}]{Ni2018}%
  \BibitemOpen
  \bibfield  {author} {\bibinfo {author} {\bibfnamefont {K.}~\bibnamefont
  {Ni}}, \bibinfo {author} {\bibfnamefont {M.}~\bibnamefont {Jerry}}, \bibinfo
  {author} {\bibfnamefont {J.~A.}\ \bibnamefont {Smith}}, \ and\ \bibinfo
  {author} {\bibfnamefont {S.}~\bibnamefont {Datta}},\ }in\ \href {\doibase
  10.1109/vlsit.2018.8510622} {\emph {\bibinfo {booktitle} {2018 {IEEE}
  Symposium on {VLSI} Technology}}}\ (\bibinfo  {publisher} {{IEEE}},\ \bibinfo
  {year} {2018})\ \bibinfo {note} {doi: 10.1109/vlsit.2018.8510622}\BibitemShut
  {NoStop}%
\bibitem [{\citenamefont {Khan}\ \emph {et~al.}(2014)\citenamefont {Khan},
  \citenamefont {Chatterjee}, \citenamefont {Wang}, \citenamefont {Drapcho},
  \citenamefont {You}, \citenamefont {Serrao}, \citenamefont {Bakaul},
  \citenamefont {Ramesh},\ and\ \citenamefont {Salahuddin}}]{Khan2014}%
  \BibitemOpen
  \bibfield  {author} {\bibinfo {author} {\bibfnamefont {A.~I.}\ \bibnamefont
  {Khan}}, \bibinfo {author} {\bibfnamefont {K.}~\bibnamefont {Chatterjee}},
  \bibinfo {author} {\bibfnamefont {B.}~\bibnamefont {Wang}}, \bibinfo {author}
  {\bibfnamefont {S.}~\bibnamefont {Drapcho}}, \bibinfo {author} {\bibfnamefont
  {L.}~\bibnamefont {You}}, \bibinfo {author} {\bibfnamefont {C.}~\bibnamefont
  {Serrao}}, \bibinfo {author} {\bibfnamefont {S.~R.}\ \bibnamefont {Bakaul}},
  \bibinfo {author} {\bibfnamefont {R.}~\bibnamefont {Ramesh}}, \ and\ \bibinfo
  {author} {\bibfnamefont {S.}~\bibnamefont {Salahuddin}},\ }\href {\doibase
  10.1038/nmat4148} {\bibfield  {journal} {\bibinfo  {journal} {Nat. Mater.}\
  }\textbf {\bibinfo {volume} {14}},\ \bibinfo {pages} {182} (\bibinfo {year}
  {2014})},\ \bibinfo {note} {doi: 10.1038/nmat4148}\BibitemShut {NoStop}%
\bibitem [{\citenamefont {Gao}\ \emph {et~al.}(2014)\citenamefont {Gao},
  \citenamefont {Khan}, \citenamefont {Marti}, \citenamefont {Nelson},
  \citenamefont {Serrao}, \citenamefont {Ravichandran}, \citenamefont
  {Ramesh},\ and\ \citenamefont {Salahuddin}}]{Gao2014}%
  \BibitemOpen
  \bibfield  {author} {\bibinfo {author} {\bibfnamefont {W.}~\bibnamefont
  {Gao}}, \bibinfo {author} {\bibfnamefont {A.}~\bibnamefont {Khan}}, \bibinfo
  {author} {\bibfnamefont {X.}~\bibnamefont {Marti}}, \bibinfo {author}
  {\bibfnamefont {C.}~\bibnamefont {Nelson}}, \bibinfo {author} {\bibfnamefont
  {C.}~\bibnamefont {Serrao}}, \bibinfo {author} {\bibfnamefont
  {J.}~\bibnamefont {Ravichandran}}, \bibinfo {author} {\bibfnamefont
  {R.}~\bibnamefont {Ramesh}}, \ and\ \bibinfo {author} {\bibfnamefont
  {S.}~\bibnamefont {Salahuddin}},\ }\href {\doibase 10.1021/nl502691u}
  {\bibfield  {journal} {\bibinfo  {journal} {Nano Lett.}\ }\textbf {\bibinfo
  {volume} {14}},\ \bibinfo {pages} {5814} (\bibinfo {year} {2014})},\ \bibinfo
  {note} {doi: 10.1021/nl502691u}\BibitemShut {NoStop}%
\bibitem [{\citenamefont {Appleby}\ \emph {et~al.}(2014)\citenamefont
  {Appleby}, \citenamefont {Ponon}, \citenamefont {Kwa}, \citenamefont {Zou},
  \citenamefont {Petrov}, \citenamefont {Wang}, \citenamefont {Alford},\ and\
  \citenamefont {O'Neill}}]{Appleby2014}%
  \BibitemOpen
  \bibfield  {author} {\bibinfo {author} {\bibfnamefont {D.~J.~R.}\
  \bibnamefont {Appleby}}, \bibinfo {author} {\bibfnamefont {N.~K.}\
  \bibnamefont {Ponon}}, \bibinfo {author} {\bibfnamefont {K.~S.~K.}\
  \bibnamefont {Kwa}}, \bibinfo {author} {\bibfnamefont {B.}~\bibnamefont
  {Zou}}, \bibinfo {author} {\bibfnamefont {P.~K.}\ \bibnamefont {Petrov}},
  \bibinfo {author} {\bibfnamefont {T.}~\bibnamefont {Wang}}, \bibinfo {author}
  {\bibfnamefont {N.~M.}\ \bibnamefont {Alford}}, \ and\ \bibinfo {author}
  {\bibfnamefont {A.}~\bibnamefont {O'Neill}},\ }\href {\doibase
  10.1021/nl5017255} {\bibfield  {journal} {\bibinfo  {journal} {Nano Lett.}\
  }\textbf {\bibinfo {volume} {14}},\ \bibinfo {pages} {3864} (\bibinfo {year}
  {2014})},\ \bibinfo {note} {doi: 10.1021/nl5017255}\BibitemShut {NoStop}%
\bibitem [{\citenamefont {Kim}\ \emph {et~al.}(2016{\natexlab{a}})\citenamefont
  {Kim}, \citenamefont {Yamada}, \citenamefont {Moon}, \citenamefont {Kwon},
  \citenamefont {An}, \citenamefont {Kim}, \citenamefont {Kim}, \citenamefont
  {Lee}, \citenamefont {Hyun}, \citenamefont {Park},\ and\ \citenamefont
  {Hwang}}]{Kim2016}%
  \BibitemOpen
  \bibfield  {author} {\bibinfo {author} {\bibfnamefont {Y.~J.}\ \bibnamefont
  {Kim}}, \bibinfo {author} {\bibfnamefont {H.}~\bibnamefont {Yamada}},
  \bibinfo {author} {\bibfnamefont {T.}~\bibnamefont {Moon}}, \bibinfo {author}
  {\bibfnamefont {Y.~J.}\ \bibnamefont {Kwon}}, \bibinfo {author}
  {\bibfnamefont {C.~H.}\ \bibnamefont {An}}, \bibinfo {author} {\bibfnamefont
  {H.~J.}\ \bibnamefont {Kim}}, \bibinfo {author} {\bibfnamefont {K.~D.}\
  \bibnamefont {Kim}}, \bibinfo {author} {\bibfnamefont {Y.~H.}\ \bibnamefont
  {Lee}}, \bibinfo {author} {\bibfnamefont {S.~D.}\ \bibnamefont {Hyun}},
  \bibinfo {author} {\bibfnamefont {M.~H.}\ \bibnamefont {Park}}, \ and\
  \bibinfo {author} {\bibfnamefont {C.~S.}\ \bibnamefont {Hwang}},\ }\href
  {\doibase 10.1021/acs.nanolett.6b01480} {\bibfield  {journal} {\bibinfo
  {journal} {Nano Letters}\ }\textbf {\bibinfo {volume} {16}},\ \bibinfo
  {pages} {4375} (\bibinfo {year} {2016}{\natexlab{a}})}\BibitemShut {NoStop}%
\bibitem [{\citenamefont {Hoffmann}\ \emph {et~al.}(2016)\citenamefont
  {Hoffmann}, \citenamefont {Pe{\v{s}}i{\'{c}}}, \citenamefont {Chatterjee},
  \citenamefont {Khan}, \citenamefont {Salahuddin}, \citenamefont {Slesazeck},
  \citenamefont {Schroeder},\ and\ \citenamefont {Mikolajick}}]{Hoffmann2016}%
  \BibitemOpen
  \bibfield  {author} {\bibinfo {author} {\bibfnamefont {M.}~\bibnamefont
  {Hoffmann}}, \bibinfo {author} {\bibfnamefont {M.}~\bibnamefont
  {Pe{\v{s}}i{\'{c}}}}, \bibinfo {author} {\bibfnamefont {K.}~\bibnamefont
  {Chatterjee}}, \bibinfo {author} {\bibfnamefont {A.~I.}\ \bibnamefont
  {Khan}}, \bibinfo {author} {\bibfnamefont {S.}~\bibnamefont {Salahuddin}},
  \bibinfo {author} {\bibfnamefont {S.}~\bibnamefont {Slesazeck}}, \bibinfo
  {author} {\bibfnamefont {U.}~\bibnamefont {Schroeder}}, \ and\ \bibinfo
  {author} {\bibfnamefont {T.}~\bibnamefont {Mikolajick}},\ }\href {\doibase
  10.1002/adfm.201602869} {\bibfield  {journal} {\bibinfo  {journal} {Adv.
  Funct. Mater.}\ }\textbf {\bibinfo {volume} {26}},\ \bibinfo {pages} {8643}
  (\bibinfo {year} {2016})},\ \bibinfo {note} {doi:
  10.1002/adfm.201602869}\BibitemShut {NoStop}%
\bibitem [{\citenamefont {Chang}\ \emph {et~al.}(2018)\citenamefont {Chang},
  \citenamefont {Avci}, \citenamefont {Nikonov}, \citenamefont {Manipatruni},\
  and\ \citenamefont {Young}}]{Chang2018}%
  \BibitemOpen
  \bibfield  {author} {\bibinfo {author} {\bibfnamefont {S.-C.}\ \bibnamefont
  {Chang}}, \bibinfo {author} {\bibfnamefont {U.~E.}\ \bibnamefont {Avci}},
  \bibinfo {author} {\bibfnamefont {D.~E.}\ \bibnamefont {Nikonov}}, \bibinfo
  {author} {\bibfnamefont {S.}~\bibnamefont {Manipatruni}}, \ and\ \bibinfo
  {author} {\bibfnamefont {I.~A.}\ \bibnamefont {Young}},\ }\href {\doibase
  10.1103/physrevapplied.9.014010} {\bibfield  {journal} {\bibinfo  {journal}
  {Phys. Rev. Appl}\ }\textbf {\bibinfo {volume} {9}} (\bibinfo {year}
  {2018}),\ 10.1103/physrevapplied.9.014010},\ \bibinfo {note} {doi:
  10.1103/physrevapplied.9.014010}\BibitemShut {NoStop}%
\bibitem [{\citenamefont {Hoffmann}\ \emph {et~al.}(2018)\citenamefont
  {Hoffmann}, \citenamefont {Max}, \citenamefont {Mittmann}, \citenamefont
  {Schroeder}, \citenamefont {Slesazeck},\ and\ \citenamefont
  {Mikolajick}}]{Hoffmann2018}%
  \BibitemOpen
  \bibfield  {author} {\bibinfo {author} {\bibfnamefont {M.}~\bibnamefont
  {Hoffmann}}, \bibinfo {author} {\bibfnamefont {B.}~\bibnamefont {Max}},
  \bibinfo {author} {\bibfnamefont {T.}~\bibnamefont {Mittmann}}, \bibinfo
  {author} {\bibfnamefont {U.}~\bibnamefont {Schroeder}}, \bibinfo {author}
  {\bibfnamefont {S.}~\bibnamefont {Slesazeck}}, \ and\ \bibinfo {author}
  {\bibfnamefont {T.}~\bibnamefont {Mikolajick}},\ }in\ \href {\doibase
  10.1109/iedm.2018.8614677} {\emph {\bibinfo {booktitle} {2018 {IEEE}
  International Electron Devices Meeting ({IEDM})}}}\ (\bibinfo  {publisher}
  {{IEEE}},\ \bibinfo {year} {2018})\BibitemShut {NoStop}%
\bibitem [{\citenamefont {Hoffmann}\ \emph {et~al.}(2019)\citenamefont
  {Hoffmann}, \citenamefont {Fengler}, \citenamefont {Herzig}, \citenamefont
  {Mittmann}, \citenamefont {Max}, \citenamefont {Schroeder}, \citenamefont
  {Negrea}, \citenamefont {Lucian}, \citenamefont {Slesazeck},\ and\
  \citenamefont {Mikolajick}}]{Hoffmann2019}%
  \BibitemOpen
  \bibfield  {author} {\bibinfo {author} {\bibfnamefont {M.}~\bibnamefont
  {Hoffmann}}, \bibinfo {author} {\bibfnamefont {F.~P.~G.}\ \bibnamefont
  {Fengler}}, \bibinfo {author} {\bibfnamefont {M.}~\bibnamefont {Herzig}},
  \bibinfo {author} {\bibfnamefont {T.}~\bibnamefont {Mittmann}}, \bibinfo
  {author} {\bibfnamefont {B.}~\bibnamefont {Max}}, \bibinfo {author}
  {\bibfnamefont {U.}~\bibnamefont {Schroeder}}, \bibinfo {author}
  {\bibfnamefont {R.}~\bibnamefont {Negrea}}, \bibinfo {author} {\bibfnamefont
  {P.}~\bibnamefont {Lucian}}, \bibinfo {author} {\bibfnamefont
  {S.}~\bibnamefont {Slesazeck}}, \ and\ \bibinfo {author} {\bibfnamefont
  {T.}~\bibnamefont {Mikolajick}},\ }\href {\doibase 10.1038/s41586-018-0854-z}
  {\bibfield  {journal} {\bibinfo  {journal} {Nature}\ }\textbf {\bibinfo
  {volume} {565}},\ \bibinfo {pages} {464} (\bibinfo {year}
  {2019})}\BibitemShut {NoStop}%
\bibitem [{\citenamefont {Kim}\ \emph {et~al.}(2019)\citenamefont {Kim},
  \citenamefont {Kim}, \citenamefont {Park}, \citenamefont {Park},
  \citenamefont {Kwon}, \citenamefont {Lee}, \citenamefont {Kim}, \citenamefont
  {Moon}, \citenamefont {Lee}, \citenamefont {Hyun}, \citenamefont {Kim},\ and\
  \citenamefont {Hwang}}]{Kim2019}%
  \BibitemOpen
  \bibfield  {author} {\bibinfo {author} {\bibfnamefont {K.~D.}\ \bibnamefont
  {Kim}}, \bibinfo {author} {\bibfnamefont {Y.~J.}\ \bibnamefont {Kim}},
  \bibinfo {author} {\bibfnamefont {M.~H.}\ \bibnamefont {Park}}, \bibinfo
  {author} {\bibfnamefont {H.~W.}\ \bibnamefont {Park}}, \bibinfo {author}
  {\bibfnamefont {Y.~J.}\ \bibnamefont {Kwon}}, \bibinfo {author}
  {\bibfnamefont {Y.~B.}\ \bibnamefont {Lee}}, \bibinfo {author} {\bibfnamefont
  {H.~J.}\ \bibnamefont {Kim}}, \bibinfo {author} {\bibfnamefont
  {T.}~\bibnamefont {Moon}}, \bibinfo {author} {\bibfnamefont {Y.~H.}\
  \bibnamefont {Lee}}, \bibinfo {author} {\bibfnamefont {S.~D.}\ \bibnamefont
  {Hyun}}, \bibinfo {author} {\bibfnamefont {B.~S.}\ \bibnamefont {Kim}}, \
  and\ \bibinfo {author} {\bibfnamefont {C.~S.}\ \bibnamefont {Hwang}},\ }\href
  {\doibase 10.1002/adfm.201808228} {\bibfield  {journal} {\bibinfo  {journal}
  {Advanced Functional Materials}\ }\textbf {\bibinfo {volume} {29}},\ \bibinfo
  {pages} {1808228} (\bibinfo {year} {2019})}\BibitemShut {NoStop}%
\bibitem [{\citenamefont {Hsu}\ \emph {et~al.}(2020)\citenamefont {Hsu},
  \citenamefont {Chang}, \citenamefont {Nikonov}, \citenamefont {Young},\ and\
  \citenamefont {Naeemi}}]{Hsu2020}%
  \BibitemOpen
  \bibfield  {author} {\bibinfo {author} {\bibfnamefont {C.-S.}\ \bibnamefont
  {Hsu}}, \bibinfo {author} {\bibfnamefont {S.-C.}\ \bibnamefont {Chang}},
  \bibinfo {author} {\bibfnamefont {D.~E.}\ \bibnamefont {Nikonov}}, \bibinfo
  {author} {\bibfnamefont {I.~A.}\ \bibnamefont {Young}}, \ and\ \bibinfo
  {author} {\bibfnamefont {A.}~\bibnamefont {Naeemi}},\ }\href {\doibase
  10.1109/ted.2020.2990891} {\bibfield  {journal} {\bibinfo  {journal} {{IEEE}
  Transactions on Electron Devices}\ }\textbf {\bibinfo {volume} {67}},\
  \bibinfo {pages} {2952} (\bibinfo {year} {2020})}\BibitemShut {NoStop}%
\bibitem [{\citenamefont {B{\"o}scke}\ \emph {et~al.}(2011)\citenamefont
  {B{\"o}scke}, \citenamefont {M{\"u}ller}, \citenamefont {Br{\"a}uhaus},
  \citenamefont {Schr{\"o}der},\ and\ \citenamefont
  {B{\"o}ttger}}]{Boescke2011}%
  \BibitemOpen
  \bibfield  {author} {\bibinfo {author} {\bibfnamefont {T.~S.}\ \bibnamefont
  {B{\"o}scke}}, \bibinfo {author} {\bibfnamefont {J.}~\bibnamefont
  {M{\"u}ller}}, \bibinfo {author} {\bibfnamefont {D.}~\bibnamefont
  {Br{\"a}uhaus}}, \bibinfo {author} {\bibfnamefont {U.}~\bibnamefont
  {Schr{\"o}der}}, \ and\ \bibinfo {author} {\bibfnamefont {U.}~\bibnamefont
  {B{\"o}ttger}},\ }\href {\doibase 10.1063/1.3634052} {\bibfield  {journal}
  {\bibinfo  {journal} {Appl. Phys. Lett.}\ }\textbf {\bibinfo {volume} {99}},\
  \bibinfo {pages} {102903} (\bibinfo {year} {2011})},\ \bibinfo {note} {doi:
  10.1063/1.3634052}\BibitemShut {NoStop}%
\bibitem [{\citenamefont {M{\"u}ller}\ \emph {et~al.}(2012)\citenamefont
  {M{\"u}ller}, \citenamefont {B{\"o}scke}, \citenamefont {Schr{\"o}der},
  \citenamefont {Mueller}, \citenamefont {Br{\"a}uhaus}, \citenamefont
  {B{\"o}ttger}, \citenamefont {Frey},\ and\ \citenamefont
  {Mikolajick}}]{Mueller2012}%
  \BibitemOpen
  \bibfield  {author} {\bibinfo {author} {\bibfnamefont {J.}~\bibnamefont
  {M{\"u}ller}}, \bibinfo {author} {\bibfnamefont {T.~S.}\ \bibnamefont
  {B{\"o}scke}}, \bibinfo {author} {\bibfnamefont {U.}~\bibnamefont
  {Schr{\"o}der}}, \bibinfo {author} {\bibfnamefont {S.}~\bibnamefont
  {Mueller}}, \bibinfo {author} {\bibfnamefont {D.}~\bibnamefont
  {Br{\"a}uhaus}}, \bibinfo {author} {\bibfnamefont {U.}~\bibnamefont
  {B{\"o}ttger}}, \bibinfo {author} {\bibfnamefont {L.}~\bibnamefont {Frey}}, \
  and\ \bibinfo {author} {\bibfnamefont {T.}~\bibnamefont {Mikolajick}},\
  }\href {\doibase 10.1021/nl302049k} {\bibfield  {journal} {\bibinfo
  {journal} {Nano Lett.}\ }\textbf {\bibinfo {volume} {12}},\ \bibinfo {pages}
  {4318} (\bibinfo {year} {2012})},\ \bibinfo {note} {doi:
  10.1021/nl302049k}\BibitemShut {NoStop}%
\bibitem [{\citenamefont {Sharma}\ \emph {et~al.}(2017)\citenamefont {Sharma},
  \citenamefont {Tapily}, \citenamefont {Saha}, \citenamefont {Zhang},
  \citenamefont {Shaughnessy}, \citenamefont {Aziz}, \citenamefont {Snider},
  \citenamefont {Gupta}, \citenamefont {Clark},\ and\ \citenamefont
  {Datta}}]{Sharma2017}%
  \BibitemOpen
  \bibfield  {author} {\bibinfo {author} {\bibfnamefont {P.}~\bibnamefont
  {Sharma}}, \bibinfo {author} {\bibfnamefont {K.}~\bibnamefont {Tapily}},
  \bibinfo {author} {\bibfnamefont {A.~K.}\ \bibnamefont {Saha}}, \bibinfo
  {author} {\bibfnamefont {J.}~\bibnamefont {Zhang}}, \bibinfo {author}
  {\bibfnamefont {A.}~\bibnamefont {Shaughnessy}}, \bibinfo {author}
  {\bibfnamefont {A.}~\bibnamefont {Aziz}}, \bibinfo {author} {\bibfnamefont
  {G.~L.}\ \bibnamefont {Snider}}, \bibinfo {author} {\bibfnamefont
  {S.}~\bibnamefont {Gupta}}, \bibinfo {author} {\bibfnamefont {R.~D.}\
  \bibnamefont {Clark}}, \ and\ \bibinfo {author} {\bibfnamefont
  {S.}~\bibnamefont {Datta}},\ }in\ \href {\doibase
  10.23919/vlsit.2017.7998160} {\emph {\bibinfo {booktitle} {2017 Symposium on
  {VLSI} Technology}}}\ (\bibinfo  {publisher} {{IEEE}},\ \bibinfo {year}
  {2017})\ \bibinfo {note} {doi: 10.23919/vlsit.2017.7998160}\BibitemShut
  {NoStop}%
\bibitem [{\citenamefont {Kobayashi}\ \emph {et~al.}(2016)\citenamefont
  {Kobayashi}, \citenamefont {Ueyama}, \citenamefont {Jang},\ and\
  \citenamefont {Hiramoto}}]{Kobayashi2016}%
  \BibitemOpen
  \bibfield  {author} {\bibinfo {author} {\bibfnamefont {M.}~\bibnamefont
  {Kobayashi}}, \bibinfo {author} {\bibfnamefont {N.}~\bibnamefont {Ueyama}},
  \bibinfo {author} {\bibfnamefont {K.}~\bibnamefont {Jang}}, \ and\ \bibinfo
  {author} {\bibfnamefont {T.}~\bibnamefont {Hiramoto}},\ }in\ \href {\doibase
  10.1109/iedm.2016.7838402} {\emph {\bibinfo {booktitle} {2016 {IEEE}
  International Electron Devices Meeting ({IEDM})}}}\ (\bibinfo  {publisher}
  {{IEEE}},\ \bibinfo {year} {2016})\ \bibinfo {note} {doi:
  10.1109/iedm.2016.7838402}\BibitemShut {NoStop}%
\bibitem [{\citenamefont {Alam}\ \emph {et~al.}(2019)\citenamefont {Alam},
  \citenamefont {Si},\ and\ \citenamefont {Ye}}]{Alam2019}%
  \BibitemOpen
  \bibfield  {author} {\bibinfo {author} {\bibfnamefont {M.~A.}\ \bibnamefont
  {Alam}}, \bibinfo {author} {\bibfnamefont {M.}~\bibnamefont {Si}}, \ and\
  \bibinfo {author} {\bibfnamefont {P.~D.}\ \bibnamefont {Ye}},\ }\href
  {\doibase 10.1063/1.5092684} {\bibfield  {journal} {\bibinfo  {journal}
  {Appl. Phys. Lett.}\ }\textbf {\bibinfo {volume} {114}},\ \bibinfo {pages}
  {090401} (\bibinfo {year} {2019})},\ \bibinfo {note} {doi:
  10.1063/1.5092684}\BibitemShut {NoStop}%
\bibitem [{\citenamefont {Liu}\ \emph {et~al.}(2020)\citenamefont {Liu},
  \citenamefont {Jiang}, \citenamefont {Ordway},\ and\ \citenamefont
  {Ma}}]{Liu2020}%
  \BibitemOpen
  \bibfield  {author} {\bibinfo {author} {\bibfnamefont {Z.}~\bibnamefont
  {Liu}}, \bibinfo {author} {\bibfnamefont {H.}~\bibnamefont {Jiang}}, \bibinfo
  {author} {\bibfnamefont {B.}~\bibnamefont {Ordway}}, \ and\ \bibinfo {author}
  {\bibfnamefont {T.~P.}\ \bibnamefont {Ma}},\ }\href {\doibase
  10.1109/led.2020.3020857} {\bibfield  {journal} {\bibinfo  {journal} {{IEEE}
  Electron Device Letters}\ }\textbf {\bibinfo {volume} {41}},\ \bibinfo
  {pages} {1492} (\bibinfo {year} {2020})}\BibitemShut {NoStop}%
\bibitem [{\citenamefont {Merz}(1956)}]{Merz1956}%
  \BibitemOpen
  \bibfield  {author} {\bibinfo {author} {\bibfnamefont {W.~J.}\ \bibnamefont
  {Merz}},\ }\href {\doibase 10.1063/1.1722518} {\bibfield  {journal} {\bibinfo
   {journal} {Journal of Applied Physics}\ }\textbf {\bibinfo {volume} {27}},\
  \bibinfo {pages} {938} (\bibinfo {year} {1956})}\BibitemShut {NoStop}%
\bibitem [{\citenamefont {Tagantsev}(2008)}]{Tagantsev2008}%
  \BibitemOpen
  \bibfield  {author} {\bibinfo {author} {\bibfnamefont {A.~K.}\ \bibnamefont
  {Tagantsev}},\ }\href {\doibase 10.1080/00150190802437746} {\bibfield
  {journal} {\bibinfo  {journal} {Ferroelectrics}\ }\textbf {\bibinfo {volume}
  {375}},\ \bibinfo {pages} {19} (\bibinfo {year} {2008})},\ \bibinfo {note}
  {doi: 10.1080/00150190802437746}\BibitemShut {NoStop}%
\bibitem [{\citenamefont {Landau}(1937)}]{Landau1937}%
  \BibitemOpen
  \bibfield  {author} {\bibinfo {author} {\bibfnamefont {L.~D.}\ \bibnamefont
  {Landau}},\ }\href@noop {} {\bibfield  {journal} {\bibinfo  {journal} {Zh.
  Eksp. Teor. Fiz}\ }\textbf {\bibinfo {volume} {7}},\ \bibinfo {pages} {19}
  (\bibinfo {year} {1937})}\BibitemShut {NoStop}%
\bibitem [{\citenamefont {Ginzburg}(1945)}]{Ginzburg1945}%
  \BibitemOpen
  \bibfield  {author} {\bibinfo {author} {\bibfnamefont {V.~L.}\ \bibnamefont
  {Ginzburg}},\ }\href@noop {} {\bibfield  {journal} {\bibinfo  {journal} {Zh.
  Eksp. Teor. Fiz}\ }\textbf {\bibinfo {volume} {15}},\ \bibinfo {pages} {739}
  (\bibinfo {year} {1945})}\BibitemShut {NoStop}%
\bibitem [{\citenamefont {Devonshire}(1949)}]{Devonshire1949}%
  \BibitemOpen
  \bibfield  {author} {\bibinfo {author} {\bibfnamefont {A.}~\bibnamefont
  {Devonshire}},\ }\href@noop {} {\bibfield  {journal} {\bibinfo  {journal}
  {The London, Edinburgh, and Dublin Philosophical Magazine and Journal of
  Science}\ }\textbf {\bibinfo {volume} {40}},\ \bibinfo {pages} {1040}
  (\bibinfo {year} {1949})},\ \bibinfo {note} {doi:
  10.1080/14786444908561372}\BibitemShut {NoStop}%
\bibitem [{\citenamefont {Si}\ \emph {et~al.}(2019)\citenamefont {Si},
  \citenamefont {Lyu},\ and\ \citenamefont {Ye}}]{Si2019}%
  \BibitemOpen
  \bibfield  {author} {\bibinfo {author} {\bibfnamefont {M.}~\bibnamefont
  {Si}}, \bibinfo {author} {\bibfnamefont {X.}~\bibnamefont {Lyu}}, \ and\
  \bibinfo {author} {\bibfnamefont {P.~D.}\ \bibnamefont {Ye}},\ }\href
  {\doibase 10.1021/acsaelm.9b00092} {\bibfield  {journal} {\bibinfo  {journal}
  {{ACS} Applied Electronic Materials}\ }\textbf {\bibinfo {volume} {1}},\
  \bibinfo {pages} {745} (\bibinfo {year} {2019})}\BibitemShut {NoStop}%
\bibitem [{\citenamefont {Lin}\ \emph {et~al.}(2005)\citenamefont {Lin},
  \citenamefont {Ye},\ and\ \citenamefont {Wilk}}]{Lin2005}%
  \BibitemOpen
  \bibfield  {author} {\bibinfo {author} {\bibfnamefont {H.~C.}\ \bibnamefont
  {Lin}}, \bibinfo {author} {\bibfnamefont {P.~D.}\ \bibnamefont {Ye}}, \ and\
  \bibinfo {author} {\bibfnamefont {G.~D.}\ \bibnamefont {Wilk}},\ }\href
  {\doibase 10.1063/1.2120904} {\bibfield  {journal} {\bibinfo  {journal}
  {Applied Physics Letters}\ }\textbf {\bibinfo {volume} {87}},\ \bibinfo
  {pages} {182904} (\bibinfo {year} {2005})}\BibitemShut {NoStop}%
\bibitem [{\citenamefont {Pe{\v{s}}i{\'{c}}}\ \emph {et~al.}(2016)\citenamefont
  {Pe{\v{s}}i{\'{c}}}, \citenamefont {Fengler}, \citenamefont {Larcher},
  \citenamefont {Padovani}, \citenamefont {Schenk}, \citenamefont {Grimley},
  \citenamefont {Sang}, \citenamefont {LeBeau}, \citenamefont {Slesazeck},
  \citenamefont {Schroeder},\ and\ \citenamefont {Mikolajick}}]{Pesic2016}%
  \BibitemOpen
  \bibfield  {author} {\bibinfo {author} {\bibfnamefont {M.}~\bibnamefont
  {Pe{\v{s}}i{\'{c}}}}, \bibinfo {author} {\bibfnamefont {F.~P.~G.}\
  \bibnamefont {Fengler}}, \bibinfo {author} {\bibfnamefont {L.}~\bibnamefont
  {Larcher}}, \bibinfo {author} {\bibfnamefont {A.}~\bibnamefont {Padovani}},
  \bibinfo {author} {\bibfnamefont {T.}~\bibnamefont {Schenk}}, \bibinfo
  {author} {\bibfnamefont {E.~D.}\ \bibnamefont {Grimley}}, \bibinfo {author}
  {\bibfnamefont {X.}~\bibnamefont {Sang}}, \bibinfo {author} {\bibfnamefont
  {J.~M.}\ \bibnamefont {LeBeau}}, \bibinfo {author} {\bibfnamefont
  {S.}~\bibnamefont {Slesazeck}}, \bibinfo {author} {\bibfnamefont
  {U.}~\bibnamefont {Schroeder}}, \ and\ \bibinfo {author} {\bibfnamefont
  {T.}~\bibnamefont {Mikolajick}},\ }\href {\doibase 10.1002/adfm.201600590}
  {\bibfield  {journal} {\bibinfo  {journal} {Advanced Functional Materials}\
  }\textbf {\bibinfo {volume} {26}},\ \bibinfo {pages} {4601} (\bibinfo {year}
  {2016})}\BibitemShut {NoStop}%
\bibitem [{\citenamefont {Kim}\ \emph {et~al.}(2016{\natexlab{b}})\citenamefont
  {Kim}, \citenamefont {Park}, \citenamefont {Kim}, \citenamefont {Lee},
  \citenamefont {Moon}, \citenamefont {Kim}, \citenamefont {Hyun},\ and\
  \citenamefont {Hwang}}]{Kim2016a}%
  \BibitemOpen
  \bibfield  {author} {\bibinfo {author} {\bibfnamefont {H.~J.}\ \bibnamefont
  {Kim}}, \bibinfo {author} {\bibfnamefont {M.~H.}\ \bibnamefont {Park}},
  \bibinfo {author} {\bibfnamefont {Y.~J.}\ \bibnamefont {Kim}}, \bibinfo
  {author} {\bibfnamefont {Y.~H.}\ \bibnamefont {Lee}}, \bibinfo {author}
  {\bibfnamefont {T.}~\bibnamefont {Moon}}, \bibinfo {author} {\bibfnamefont
  {K.~D.}\ \bibnamefont {Kim}}, \bibinfo {author} {\bibfnamefont {S.~D.}\
  \bibnamefont {Hyun}}, \ and\ \bibinfo {author} {\bibfnamefont {C.~S.}\
  \bibnamefont {Hwang}},\ }\href {\doibase 10.1039/c5nr05339k} {\bibfield
  {journal} {\bibinfo  {journal} {Nanoscale}\ }\textbf {\bibinfo {volume}
  {8}},\ \bibinfo {pages} {1383} (\bibinfo {year}
  {2016}{\natexlab{b}})}\BibitemShut {NoStop}%
\bibitem [{\citenamefont {Chouprik}\ \emph {et~al.}(2019)\citenamefont
  {Chouprik}, \citenamefont {Spiridonov}, \citenamefont {Zarubin},
  \citenamefont {Kirtaev}, \citenamefont {Mikheev}, \citenamefont
  {Lebedinskii}, \citenamefont {Zakharchenko},\ and\ \citenamefont
  {Negrov}}]{Chouprik2019}%
  \BibitemOpen
  \bibfield  {author} {\bibinfo {author} {\bibfnamefont {A.}~\bibnamefont
  {Chouprik}}, \bibinfo {author} {\bibfnamefont {M.}~\bibnamefont
  {Spiridonov}}, \bibinfo {author} {\bibfnamefont {S.}~\bibnamefont {Zarubin}},
  \bibinfo {author} {\bibfnamefont {R.}~\bibnamefont {Kirtaev}}, \bibinfo
  {author} {\bibfnamefont {V.}~\bibnamefont {Mikheev}}, \bibinfo {author}
  {\bibfnamefont {Y.}~\bibnamefont {Lebedinskii}}, \bibinfo {author}
  {\bibfnamefont {S.}~\bibnamefont {Zakharchenko}}, \ and\ \bibinfo {author}
  {\bibfnamefont {D.}~\bibnamefont {Negrov}},\ }\href {\doibase
  10.1021/acsaelm.8b00046} {\bibfield  {journal} {\bibinfo  {journal} {{ACS}
  Applied Electronic Materials}\ }\textbf {\bibinfo {volume} {1}},\ \bibinfo
  {pages} {275} (\bibinfo {year} {2019})}\BibitemShut {NoStop}%
\bibitem [{\citenamefont {Goh}\ \emph {et~al.}(2020)\citenamefont {Goh},
  \citenamefont {Cho}, \citenamefont {Park},\ and\ \citenamefont
  {Jeon}}]{Goh2020}%
  \BibitemOpen
  \bibfield  {author} {\bibinfo {author} {\bibfnamefont {Y.}~\bibnamefont
  {Goh}}, \bibinfo {author} {\bibfnamefont {S.~H.}\ \bibnamefont {Cho}},
  \bibinfo {author} {\bibfnamefont {S.-H.~K.}\ \bibnamefont {Park}}, \ and\
  \bibinfo {author} {\bibfnamefont {S.}~\bibnamefont {Jeon}},\ }\href {\doibase
  10.1039/d0nr00933d} {\bibfield  {journal} {\bibinfo  {journal} {Nanoscale}\
  }\textbf {\bibinfo {volume} {12}},\ \bibinfo {pages} {9024} (\bibinfo {year}
  {2020})}\BibitemShut {NoStop}%
\bibitem [{\citenamefont {Lomenzo}\ \emph {et~al.}(2020)\citenamefont
  {Lomenzo}, \citenamefont {Richter}, \citenamefont {Mikolajick},\ and\
  \citenamefont {Schroeder}}]{Lomenzo2020}%
  \BibitemOpen
  \bibfield  {author} {\bibinfo {author} {\bibfnamefont {P.~D.}\ \bibnamefont
  {Lomenzo}}, \bibinfo {author} {\bibfnamefont {C.}~\bibnamefont {Richter}},
  \bibinfo {author} {\bibfnamefont {T.}~\bibnamefont {Mikolajick}}, \ and\
  \bibinfo {author} {\bibfnamefont {U.}~\bibnamefont {Schroeder}},\ }\href
  {\doibase 10.1021/acsaelm.0c00184} {\bibfield  {journal} {\bibinfo  {journal}
  {{ACS} Applied Electronic Materials}\ }\textbf {\bibinfo {volume} {2}},\
  \bibinfo {pages} {1583} (\bibinfo {year} {2020})}\BibitemShut {NoStop}%
\bibitem [{\citenamefont {Jo}\ \emph {et~al.}(2009)\citenamefont {Jo},
  \citenamefont {Yang}, \citenamefont {Kim}, \citenamefont {Lee}, \citenamefont
  {Yoon}, \citenamefont {Park}, \citenamefont {Jo}, \citenamefont {Jung},\ and\
  \citenamefont {Noh}}]{Jo2009}%
  \BibitemOpen
  \bibfield  {author} {\bibinfo {author} {\bibfnamefont {J.~Y.}\ \bibnamefont
  {Jo}}, \bibinfo {author} {\bibfnamefont {S.~M.}\ \bibnamefont {Yang}},
  \bibinfo {author} {\bibfnamefont {T.~H.}\ \bibnamefont {Kim}}, \bibinfo
  {author} {\bibfnamefont {H.~N.}\ \bibnamefont {Lee}}, \bibinfo {author}
  {\bibfnamefont {J.-G.}\ \bibnamefont {Yoon}}, \bibinfo {author}
  {\bibfnamefont {S.}~\bibnamefont {Park}}, \bibinfo {author} {\bibfnamefont
  {Y.}~\bibnamefont {Jo}}, \bibinfo {author} {\bibfnamefont {M.~H.}\
  \bibnamefont {Jung}}, \ and\ \bibinfo {author} {\bibfnamefont {T.~W.}\
  \bibnamefont {Noh}},\ }\href {\doibase 10.1103/physrevlett.102.045701}
  {\bibfield  {journal} {\bibinfo  {journal} {Physical Review Letters}\
  }\textbf {\bibinfo {volume} {102}} (\bibinfo {year} {2009}),\
  10.1103/physrevlett.102.045701}\BibitemShut {NoStop}%
\bibitem [{\citenamefont {Khan}\ \emph {et~al.}(2016)\citenamefont {Khan},
  \citenamefont {Radhakrishna}, \citenamefont {Chatterjee}, \citenamefont
  {Salahuddin},\ and\ \citenamefont {Antoniadis}}]{Khan2016}%
  \BibitemOpen
  \bibfield  {author} {\bibinfo {author} {\bibfnamefont {A.~I.}\ \bibnamefont
  {Khan}}, \bibinfo {author} {\bibfnamefont {U.}~\bibnamefont {Radhakrishna}},
  \bibinfo {author} {\bibfnamefont {K.}~\bibnamefont {Chatterjee}}, \bibinfo
  {author} {\bibfnamefont {S.}~\bibnamefont {Salahuddin}}, \ and\ \bibinfo
  {author} {\bibfnamefont {D.~A.}\ \bibnamefont {Antoniadis}},\ }\href
  {\doibase 10.1109/ted.2016.2612656} {\bibfield  {journal} {\bibinfo
  {journal} {{IEEE} Transactions on Electron Devices}\ }\textbf {\bibinfo
  {volume} {63}},\ \bibinfo {pages} {4416} (\bibinfo {year}
  {2016})}\BibitemShut {NoStop}%
\bibitem [{\citenamefont {Zhou}\ \emph {et~al.}(2019)\citenamefont {Zhou},
  \citenamefont {Han}, \citenamefont {Xu}, \citenamefont {Li}, \citenamefont
  {Peng}, \citenamefont {Liu}, \citenamefont {Zhang}, \citenamefont {Sun},
  \citenamefont {Zhang},\ and\ \citenamefont {Hao}}]{Zhou2019}%
  \BibitemOpen
  \bibfield  {author} {\bibinfo {author} {\bibfnamefont {J.}~\bibnamefont
  {Zhou}}, \bibinfo {author} {\bibfnamefont {G.}~\bibnamefont {Han}}, \bibinfo
  {author} {\bibfnamefont {N.}~\bibnamefont {Xu}}, \bibinfo {author}
  {\bibfnamefont {J.}~\bibnamefont {Li}}, \bibinfo {author} {\bibfnamefont
  {Y.}~\bibnamefont {Peng}}, \bibinfo {author} {\bibfnamefont {Y.}~\bibnamefont
  {Liu}}, \bibinfo {author} {\bibfnamefont {J.}~\bibnamefont {Zhang}}, \bibinfo
  {author} {\bibfnamefont {Q.-Q.}\ \bibnamefont {Sun}}, \bibinfo {author}
  {\bibfnamefont {D.~W.}\ \bibnamefont {Zhang}}, \ and\ \bibinfo {author}
  {\bibfnamefont {Y.}~\bibnamefont {Hao}},\ }\href {\doibase
  10.1109/led.2018.2886426} {\bibfield  {journal} {\bibinfo  {journal} {{IEEE}
  Electron Device Letters}\ }\textbf {\bibinfo {volume} {40}},\ \bibinfo
  {pages} {329} (\bibinfo {year} {2019})}\BibitemShut {NoStop}%
\end{thebibliography}%

\end{document}